\newcommand{\eqn}[1]{Eq.~(\ref{#1})}

\def\simge{
    \mathrel{\rlap{\raise 0.511ex
        \hbox{$>$}}{\lower 0.511ex \hbox{$\sim$}}}}

\documentclass[runningheads]{svmult}

\usepackage{makeidx}   
\usepackage{graphicx}  
\usepackage{epsfig}
\usepackage{subeqnar}  
\usepackage{multicol}  
\usepackage{physprbb}  
\makeindex             

\newcommand{\be}{\begin{equation}}
\newcommand{\ee}{\end{equation}}

\begin{document}
\title*{Baryon Spectroscopy in Lattice QCD}
\toctitle{Baryon Spectroscopy in Lattice QCD}
%
%
\titlerunning{Baryon Spectroscopy in Lattice QCD}
%
\author{D. B.~Leinweber\inst{1}
\and W.~Melnitchouk\inst{2}
\and D. G.~Richards\inst{2}
\and A. G.~Williams\inst{1}
\and J. M.~Zanotti\inst{3}}
\authorrunning{D. B.~Leinweber {\em et al.}}
%
%
\institute{Department of Physics and Mathematical Physics and\\
        Special Research Centre for the
        Subatomic Structure of Matter,                       \\
        University of Adelaide, 5005, Australia
\and
        Jefferson Lab, 12000 Jefferson Avenue,
        Newport News, VA 23606, USA
\and John von Neumann-Institut f\"ur Computing
  NIC, \\
  Deutsches Elektronen-Synchrotron DESY, D-15738 Zeuthen, Germany}

\maketitle              

\vspace{-7cm}
\null \hfill ADP-04-10/T592 \\
\null \hfill JLAB-THY-04-223 \\
\null \hfill DESY-04-078 \\
\vspace{-20pt}
\vspace{-20pt}
\vspace{-20pt}
\vspace{7cm}

\begin{abstract}
We review recent developments in the study of excited baryon
spectroscopy in lattice QCD.
After introducing the basic methods used to extract masses
from correlation functions, we discuss various interpolating fields
and lattice actions commonly used in the literature.
We present a survey of results of recent calculations of excited
baryons in quenched QCD, and outline possible future directions
in the study of baryon spectra.
\end{abstract}

\section{Introduction and Motivation}

One of the primary tools used for studying the forces which confine
quarks inside hadrons, and determining the relevant effective degrees
of freedom in strongly coupled QCD, has been baryon and meson
spectroscopy.
This is a driving force behind the current experimental $N^*$ programme
at Jefferson Lab, which is accumulating data of unprecedented quality
and quantity on various $N \to N^*$ transitions \cite{CLAS}.
The prospects of studying mesonic spectra, and in particular the role
played by gluonic excitations, has been a major motivation for future
facilities such as CLEO-c \cite{CLEOC}, the anti-proton facility at
GSI (PANDA) \cite{PANDA}, and the Hall~D programme at a 12~GeV energy 
upgraded CEBAF \cite{HALLD}.
With the increased precision of the new $N^*$ data comes a growing
need to understand the observed spectrum within QCD.  QCD-inspired
phenomenological models, whilst successful in describing many features
of the $N^*$ spectrum \cite{CR}, leave many questions unanswered.
%

One of the long-standing puzzles in baryon spectroscopy has been the
low mass of the first positive parity excitation of the nucleon,
the $N^*(1440)$ Roper resonance, compared with the lowest lying odd
parity excitation.
In valence quark models with harmonic oscillator potentials, the
$J^P = {1\over 2}^-$ state naturally occurs below the $N=2$,
${1\over 2}^+$ excitation \cite{IK}.
Without fine tuning of parameters, valence quark models tend to leave
the Roper mass too high.
Similar difficulties in the level orderings appear for the
$J^P = {3\over 2}^+$ $\Delta^*(1600)$ and ${1\over 2}^+$ 
$\Sigma^*(1690)$ resonances, which have led to speculations that the 
Roper resonances may be more appropriately viewed as hybrid baryon
states with explicitly excited glue field configurations \cite{LBL},
rather than as radial excitations of a three-quark configuration.
Other scenarios portray the Roper resonances as ``breathing modes'' 
of the ground states \cite{BREATHE}, or states which can be described
in terms of meson-baryon dynamics alone \cite{FZJ}.

Another challenge for spectroscopy is presented by the anomalously
small mass of the odd parity $\Lambda(1405)$ hyperon.
This is alternatively interpreted as a true three-quark state, or as
a hadronic molecule arising from strong coupled channel effects
between $\Sigma\pi$, $\bar K N$, $\ldots$ states \cite{L1405}.
In fact, the role played by Goldstone bosons in baryon spectroscopy
has received considerable attention recently
\cite{MASSEXTR,YOUNG,Morel:2003fj}.

Mass splittings between states within SU(3) quark-model multiplets
provide another important motivation for studying the baryon spectrum.
The dynamical origin of hyperfine splittings in quark models has
traditionally been attributed to the colour-magnetic one gluon
exchange mechanism \cite{RUJ}.
On the other hand, there have been attempts recently to explain the
hyperfine splittings and level orderings in terms of a spin-flavour
interaction associated with the exchange of a pseudoscalar nonet of
Goldstone bosons between quarks \cite{GBE}.
Understanding the mass splitting between the $J^P={1\over 2}^-$
$N^*(1535)$ and ${3\over 2}^-$ $N^*(1520)$, or between the
${1\over 2}^-$ $\Delta^*(1620)$ and ${3\over 2}^-$ $\Delta^*(1700)$,
can help identify the important mechanisms associated with the
hyperfine interactions, and shed light on the spin-orbit force,
which has been a central mystery in spectroscopy \cite{SUMRULES}.
In valence quark models, the degeneracy between the $N{1\over 2}^-$
and $N{3\over 2}^-$ can be broken by a tensor force associated with
mixing between the $N^2$ and $N^4$ representations of SU(3) \cite{CR}, 
although this generally leaves the $N{3\over 2}^-$ at a higher energy 
than the $N{1\over 2}^-$.
In contrast, a spin-orbit force is necessary to split the
$\Delta{3\over 2}^-$ and $\Delta{1\over 2}^-$ states.
In the Goldstone boson exchange model \cite{GBE}, both of these
pairs of states are degenerate.

Similarly, neither spin-flavour nor colour-magnetic interactions is able
to account for the mass splitting between the $\Lambda(1405)$ and the
$J^P = {3\over 2}^-$ $\Lambda^*(1520)$.
A splitting between these can arise in constituent quark models with
a spin-orbit interaction, but the required strength of the
interaction leads to spurious mass splittings elsewhere \cite{CR,CI}.
In fact, model-independent analyses of baryons in the large $N_c$
limit \cite{LARGE_NC} have found that these mass splittings receive
important contributions from operators that do not have a simple quark
model interpretation \cite{LARGE_NC}, such as those simultaneously
coupling spin, isospin and orbital angular momentum, as anticipated
by early QCD sum-rule analyses \cite{SUMRULES}.
Of course, the coefficients of the various operators in such an
analysis must be determined phenomenologically and guidance from
lattice QCD is essential.

The large number of ``missing'' resonances, predicted by the
constituent quark model and its generalisations but not observed
experimentally, presents another problem for spectroscopy.
If these states do not exist, this may suggest that perhaps a
quark--diquark picture (with fewer degrees of freedom) could afford
a more efficient description.
Alternatively, the missing states could simply have weak couplings
to the $\pi N$ system \cite{CR}.
%

Apart from these long-standing puzzles, interest in baryon spectroscopy
has been further stimulated with the recent discovery of the exotic
strangeness $S=+1$ $\Theta^+$ pentaquark \cite{5Q}, which has a minimum
5 valence quark ($uudd\bar s$) content.
Postulated long ago, the $\Theta^+$ appears to have eluded searches
until now, and understanding its properties and internal structure
has become a major challenge for spectroscopy.

A consistent description of the baryon spectrum within low-energy
models requires a satisfactory resolution of the old questions of
spectroscopy, as well as insight into the new puzzles.
On both of these fronts real progress may only come with the help of
first-principle calculations of the spectrum in lattice QCD, which at
present is the only method capable of determining hadron properties
directly from the fundamental quark and gluon theory.
Considerable progress has been made in the past few years in
calculating hadronic properties in both quenched and full QCD, with
the ground state masses now understood at the few percent level.

Compared with simulations of hadron ground state properties, however,
the calculation of the excited nucleon spectrum places particularly
heavy demands on lattice spectroscopy.
The excited nucleon states are expected to be large, with the size of
a state expected to double with each increase in orbital angular
momentum.
Lattice studies of the excited nucleon spectrum therefore require
large lattice volumes, with correspondingly large computational
resources.
Furthermore, the states are relatively massive, requiring a fine 
lattice spacing, at least in the temporal direction.
Recent advances in computational capabilities and more efficient
algorithms have enabled the first dedicated lattice QCD simulations
of the excited states of the nucleon to be undertaken.

Of course, the calculation of the hadronic spectrum faces a formidable
challenge in describing excited state decays, which is relevant in both
full and quenched QCD.
Lattice studies of excited hadrons are possible because at the
current unphysically large quark masses and finite volumes used
in the simulations many excited states are stable \cite{RHO}.
Contact with experiment can be made via extrapolations incorporating
the nonanalytic behaviour predicted by chiral effective field theory.


The rest of this review is laid out as follows.
After briefly reviewing in Section~2 the history of lattice calculations
of the excited baryon spectrum, in Section~3 we outline the basic
lattice techniques for extracting masses from correlation functions,
including variational methods and Bayesian statistics.
Section~4 describes the construction of a suitable basis of
interpolating fields for any baryon we may wish to investigate.
Section~5 outlines the simplest interpolating fields for spin-1/2
and spin-3/2 baryons, and how these can be used in the construction
of lattice correlation functions.
In Section~6, we present a survey of recent $N^*$ results for both
positive and negative parity baryons.
Finally, in Sec.~7 we discuss conclusions of the existing studies,
and speculate on future directions for the study of baryon spectra.
Some technical aspects of the correlation matrix formalism for
calculating masses, coupling strengths and optimal interpolating
fields are described in the Appendix.

\section{History of Lattice $N^*$ Calculations}
\label{history}

The history of excited baryons on the lattice is rather short,
although recently there has been growing interest in finding new
techniques to isolate excited baryons, motivated partly by the
experimental $N^*$ programme at Jefferson Lab.  An attempt to
determine the first positive-parity excitation of the nucleon using
variational methods was made by the UKQCD collaboration\cite{ukqcd93}.
A more detailed analysis of the positive parity excitation of the
nucleon was performed by Leinweber \cite{DEREK} using Wilson fermions
and an operator product expansion spectral ansatz.  DeGrand and Hecht
\cite{DEGRAND92} used a wave function ansatz to access $P$-wave
baryons, with Wilson fermions and relatively heavy quarks.
Subsequently, Lee and Leinweber \cite{LL} introduced a parity
projection technique to study the negative parity ${1 \over 2}^-$
states using an ${\cal O}(a^2)$ tree-level tadpole-improved D$_{\chi
34}$ quark action, and an ${\cal O}(a^2)$ tree-level tadpole-improved
gauge action.  Following this, Lee \cite{LEE} reported results using a
D$_{234}$ quark action with an improved gauge action on an anisotropic
lattice to study the ${1 \over 2}^+$ and ${1 \over 2}^-$ excitations
of the nucleon.
An anisotropic lattice with an ${\cal O}(a)$ improved quark action
was also used by Nakajima {\it et al.} \cite{Nakajima:2001js} to study
excited states of octet and decuplet baryons.

The RIKEN-BNL group \cite{BNL} has stressed the importance of
maintaining chiral symmetry on the lattice.
At finite lattice spacing the Wilson fermion action is known to
explicitly break the chiral symmetry of continuum QCD.
A solution to this problem is provided through the introduction of a
fifth dimension, which allows chiral symmetry to be maintained even at
non-zero lattice spacing \cite{DWF}.
The resulting domain wall fermion action was used in Ref.~\cite{BNL}
to compute the masses of the $N{1\over 2}^-$ and $N{1\over 2}^+$
excited states.
The analysis has recently been extended by studying the finite volume
effects of the first $N{1\over 2}^+$ excited state extracted using
maximum entropy methods \cite{Sasaki:2003xc}.

A nonperturbatively ${\cal O}(a)$-improved Sheikholeslami-Wohlert (SW)
\cite{CLOVER}, or clover, fermion action was used by Richards
{\it et al.} \cite{RICHARDS} to study the $N{1\over 2}^-$ and
$\Delta{3\over 2}^-$ states.
By appropriately choosing the coefficient of the improvement term, all
${\cal O}(a)$ discretisation uncertainties can be removed, ensuring
that continuum-like results are obtained at a finite lattice spacing.
For more details on ${\cal O}(a)$-improvement, see the contribution by
Zanotti {\it et al.} in this lecture series.

The BGR collaboration \cite{FP} has been investigating the masses of
the $N{1\over 2}^-$ and $N{1\over 2}^+$ excited states calculated with
chirally improved (CI) and Fixed Point (FP) fermions. Both of these
actions offer the advantage of improved chiral properties over
traditional Wilson-style fermions, without the cost associated with
Ginsparg-Wilson fermions, which possess an exact analogue of chiral
symmetry.

The CSSM Lattice Collaboration has used an ${\cal O}(a^2)$ improved gluon
action and the ${\cal O}(a)$-improved Fat Link Irrelevant Clover
(FLIC) fermion action 
\cite{FATJAMES} to perform a comprehensive study of the spectrum of
positive and negative parity baryons \cite{CSSM,CSSM32}.
Excited state masses in both the octet and decuplet multiplets have
been computed, including the $N{1\over 2}^\pm$, $N{3\over 2}^\pm$,
$\Sigma{1\over 2}^\pm$, $\Lambda{1\over 2}^\pm$, $\Xi{1\over 2}^\pm$,
$\Delta{1\over 2}^\pm$ and $\Delta{3\over 2}^\pm$ baryons.
The formulation of FLIC fermions and a review of its scaling and light
quark mass phenomenology is reviewed by Zanotti {\it et al.} in this
lecture series. 

Constrained-fitting methods originating from Bayesian priors have also
recently been used by Lee {\em et al.} \cite{BAYESIAN} to study the two
lowest octet and decuplet positive and negative parity baryons, using
overlap fermions with pion masses down to $\sim 180$~MeV.
These authors have addressed the difficult issue of contamination of the
first excited state of the nucleon with quenched $\eta'N$ artifacts
\cite{Dong:2003zf}.
While these authors claim to have observed the Roper in quenched QCD,
it remains to be demonstrated that this conclusion is independent of 
the Bayesian-prior assumed in their analysis \cite{DEREK}.
It would be interesting to examine these correlation functions using
correlation matrix techniques or alternative Bayesian techniques such
as the Maximum Entropy Method.

As mentioned in Sec.~1, we are currently seeing intense interest
in exotic pentaquark spectroscopy, and recently the first lattice 
studies of 5-quark states have appeared
\cite{5Q_CSIKOR,5Q_SASAKI,5Q_CHIU}.
Here the unique advantage of lattice QCD can come to the fore, in
predicting, for instance, the parity and spin of the lowest lying
pentaquark state, which are as yet undetermined experimentally.
The first two of these early studies have used the standard Wilson
gauge and fermion action, on lattice sizes of $L=1.2$--2.7~fm
\cite{5Q_CSIKOR}, and $L=2.2$~fm \cite{5Q_SASAKI}, while the third
used domain wall fermions with a Wilson gauge action on a
$L=1.8$~fm lattice.

The main challenge here has been the construction of lattice operators
for states with 4 quarks and an antiquark, which can be variously
constructed in terms of ($qqq$)($q\bar s$) ``nucleon--meson''
\cite{5Q_CSIKOR} or ($qq$)($qq$)$\bar s$ ``diquark-diquark-$\bar s$''
\cite{5Q_SASAKI,5Q_CHIU} operators.
While two of these studies \cite{5Q_CSIKOR,5Q_SASAKI} appear to favor
negative parity for the lowest lying pentaquark state, it is not clear
that this state is a new resonant combination of $K\, N$ in $S$ wave.
The most recent study \cite{5Q_CHIU} reports an even parity ground
state, and moreover identifies the Roper resonance as a pentaquark
state.
The relatively small volumes, limited set of interpolating fields,
and naive linear extrapolations means that these results must be
regarded as exploratory at present.
New studies using the FLIC fermion action with several different
interpolating fields are currently in progress \cite{5Q_FLIC}.

We are clearly witnessing an exciting period in which the field of
baryon spectroscopy on the lattice is beginning to flourish.
%
\section{Lattice Techniques}
\subsection{Spectroscopy Recipe}

The computation of the spectrum of states in lattice QCD is in
principle straightforward.
The building blocks are the quark propagators
\begin{equation}
S^{ij}_{\alpha\beta}(x,y) = \langle 0 | \psi^i_{\alpha}(x)
\bar{\psi}^j_{\beta} (y) | 0 \rangle,
\label{eq:quark_prop}
\end{equation}
computed on an ensemble of gauge configurations, composed of
gauge field links $U_{\mu}(x) \simeq \exp \{i a g A_\mu(x)\}$.  The
calculation then proceeds 
as follows:
\begin{enumerate}
\item Choose an interpolating operator ${\cal O}$ that has a good
overlap with $P$, the state of interest,
\[
\langle 0\mid {\cal O} \mid P\rangle \ne 0,
\]
and ideally a small overlap with other states having the same quantum
numbers.
\item Form the time-sliced correlation function
\[
G(t,\vec{p}) = \sum_{\vec{x}} \langle {\cal O}(\vec{x}, t) {\cal O}^{\dagger}
(\vec{0}, 0) \rangle e^{-i \vec{p}\cdot\vec{x}},
\label{eq:corrs_cons}
\]
which can be expressed by a Wick expansion in terms of the elemental
quark propagators of \eqn{eq:quark_prop}.
\item  Insert a complete set of states between ${\cal O}$ and ${\cal
O}^{\dagger}$.  The time-sliced
sum puts the intermediate states at definite momentum, and one finds 
\begin{eqnarray*}
G(t,\vec{p}) & = & \sum_{\vec{x}} \sum_P \int \frac{d^3k}{(2 \pi)^3 2 E(\vec{k})}
e^{-i \vec{p} \cdot \vec{x}} \langle 0 |
{\cal O}(\vec{x}, t) | P(\vec{k}) \rangle \langle P(\vec{k}) | {\cal
O}^{\dagger} (\vec{0},0) | 0 \rangle\\
& = & \sum_P \frac{\mid \langle 0 \mid {\cal O} \mid P
\rangle \mid^2}{2 \, E_P(\vec{p}) } e^{-i E_P(\vec{p}) t}\ ,
\end{eqnarray*}
where the sum over $P$ includes the contributions from two-particle
and higher states.
\item Continue to Euclidean space $t \rightarrow -i t$, giving
\begin{equation}
G(t,\vec p) = \sum_P \frac{\mid \langle 0 \mid {\cal O} \mid P
\rangle \mid^2}{2 \, E_P(\vec{p})} e^{- E_P(\vec{p}) t}.
\label{eq:spectral_sum}
\end{equation}
\end{enumerate}

At large Euclidean times, the lightest state dominates the spectral sum in
Eq.~(\ref{eq:spectral_sum}), and we can extract the ground state mass.
The determination of this ground state mass for the states of lowest
spin has been the benchmark calculation of lattice QCD since its
inception.  However, our goal is to build up a more complete
description of the baryon spectrum, ultimately determining not only
the masses of some of the higher spin particles, but also the masses
of the radial excitations.  In the remainder of this section, we will
address two issues that are crucial to attaining this goal: the
application of variational techniques 
to isolate the higher excitations in~\eqn{eq:spectral_sum},
and the use of the appropriate statistical fitting techniques to
reliably extract the energies of those excitations.
In the following section, we will describe the construction of the
nucleon interpolating operators.

\subsection{Fitting Techniques}

\subsubsection{Variational Methods}

To confidently extract other than the lowest-lying state in
\eqn{eq:spectral_sum}, it is crucial to have more than a single
interpolating operator ${\cal O}_i$ in order to appeal to
variational methods to determine the spectrum of
states \cite{michael85,lw90}.  Our aim is to measure every element of
the matrix of correlators
\begin{equation}
 G_{ij}(t) = \langle \sum_{\vec{x}} {\cal O}_i(\vec{x},t) {\cal
 O}^{\dagger}_j(0) \rangle \label{eq:matrix_corr}.
\end{equation}
We now consider the eigenvalue equation
\begin{equation}
G(t) u = \lambda(t,t_0) G(t_0) u,\label{eq:corr_ev}
\end{equation}
for some eigenfunction $u$ of $G^{-1}( t_0 ) G(t)$, with $t_0$ fixed.
For the sake of illustration, we will consider a simple system with
only two independent states, and a $2 \times 2$ matrix of correlators.
Then the eigenvalues of \eqn{eq:corr_ev}
satisfy
\begin{eqnarray}
\lambda_+(t,t_0) & = & e^{- (t - t_0) E_0}\ , \nonumber\\
\lambda_-(t,t_0) & = & e^{- (t - t_0) E_1}\ ,
\end{eqnarray}
and we have an exact separation between the energies of the two
states, with coefficients growing exponentially with $t_0$.  For the
physical case of more than two states, there are exponential
corrections arising from the states of higher energy.  Ideally, to
suppress the contribution of these higher energy states we would wish
to choose $t_0$ to have as large a value as possible.  However,
increasing $t_0$ comes at the price of increasing statistical noise,
and therefore we are generally obliged to take $t_0$ close to the
origin.

In practice, it is usual to adopt a variation of the above method, and
not to attempt to diagonalise the transfer matrix at each time slice,
but rather to choose a matrix of eigenvectors $V(t_0)$ for some $t_0$ close to
the source that diagonalises $G(t_0)^{-1} G(t_0 + 1)$ .
For the case of a $2 \times 2$ matrix with only two states, the
diagonal elements of the matrix
\[
V(t_0+1)^{-1} G(t_0)^{-1} G(t) V(t_0 + 1)
\]
are indeed equivalent to the
eigenvalues of $G(t_0)^{-1} G(t_0 + 1)$. A more pedagogical discussion of these
concepts can be found in the Appendix.

The efficacy of this method relies on having a basis of correlators
that can delineate the structure of the first few states, together
with sufficiently high-quality statistics that the elements of the
correlator matrix can be well determined.  In the case of the glueball
spectrum, the computational cost is dominated by the cost of
generating the gauge configurations, and the overhead of measuring
extra correlators is negligible.  Furthermore, we are able to improve
the statistical quality of the data by using translational
invariance to average over the position of the source coordinate in
\eqn{eq:corrs_cons}.  Variational techniques have been essential, and
very successful, in the extraction of the glueball spectrum
\cite{MCNEILE,Morningstar:1999rf}.

In the case of operators containing quark fields, there is generally a
considerable computational cost associated with the measurement of
additional correlators.  We will discuss the construction of such
operators in the next section, but it is straightforward to see how
this cost arises.  Recall that the basic building blocks of
spectroscopy are the quark propagators of \eqn{eq:quark_prop}, which
satisfy
\begin{equation}
M^{ij}_{\alpha\beta}(x,y) S^{jk}_{\beta\gamma}(y,z) = \delta^{ik}
\delta_{\alpha \gamma} \delta_{xz}.
\label{eq:matrix_inverse}
\end{equation}
where $M$ is the fermion matrix of the fermion action $\overline \psi
\, M \, \psi$.
The propagator is usually obtained by standard sparse matrix inversion
methods, such as conjugate gradient.  These methods are applicable to
a fixed source vector on the right hand side of
\eqn{eq:matrix_inverse}, and additional correlators require the
inversion of $M$ for additional source vectors.  A corollary of this
is that the matrix of correlators is obtained for fixed source
coordinate, and therefore we cannot in general use translational
invariance to average over the source coordinate in the manner of
glueball calculations.

\subsubsection{Bayesian Statistics}

Given the computational cost of measuring additional correlators, it
is important to extract the greatest possible information from the
those that are measured.  Furthermore, it is vital that the fitting
procedure be as reliable as possible, in particular by ensuring that
the extracted masses are not adversely affected by contamination from
higher excitations.  Historically, lattice calculations employed
\textit{maximum likelyhood} fits to correlators such as 
\eqn{eq:spectral_sum}, using a single or at most two states in the
spectral sum.  An acceptable $\chi^2/{\rm d.o.f.}$ would only be
obtained if the fitted data were restricted to large temporal
separations, in which one or two states did indeed dominate the
spectral sum.  Thus the fits ignored the data at small temporal
separations, which we have already seen has the largest
signal-to-noise ratio.

It is natural to ask whether one can extract useful information from
the data at small values of $t$, and a means of so doing is provided
through the use of Bayesian Statistics.
We will return to a further discussion of the the Bayesian
approach in our survey of results.

%

\section{Interpolating Fields}

The variational techniques and fitting methods described above require
a suitable basis of interpolating fields from which operators can be
constructed which mimic the structure of each of the states to be
extracted.
In order to do so, we firstly have to consider the possible quantum
numbers by which states are classified, and the extent to which they
remain good quantum numbers in lattice calculations.
We then have to consider the spatial structure of the states, and in
particular their spatial extents, and whether we can construct
interpolating operators that reflect the structure, through the use of,
say, smeared interpolating fields.
We will begin this section by detailing the quantum numbers with which
states are classified, namely flavour and parity, and the degree to
which they are good quantum numbers on the lattice.
We will then proceed to discuss the use of spatially extended
interpolating operators, known
as smearing.  Finally, we will describe the operators that have been
employed thus far in lattice calculations, and discuss these in terms
of continuum quark model operators.

\subsection{Continuum and Lattice Symmetries}

Baryon states are classified by their flavour structure, either
according to $SU(2)$ (isospin), or $SU(3)$ (strangeness), parity, and
total spin.  In nearly all lattice calculations of the hadron
spectrum, exact isospin symmetry is imposed so the $m_u = m_d$, and
electromagnetic effects are ignored.  Thus the flavour
structure of baryon states composed of light ($u,d,s$) quarks is
specified according to \textit{Total Isospin} $I$, $I_3$, and the
\textit{strangeness} $S$, and the naming of baryon states follows from
this labeling, as detailed in Table~\ref{tab:flavour}.
\begin{table}
\begin{center}
\caption{The table shows the flavour classification of baryon states
  constructed from light ($u,d,s$) quarks, together with a representative
  flavour structure in terms of three valence quarks for the case
  $I_3 = I$.}
\label{tab:flavour}
\begin{tabular}{r|r|c|c}
$I$ & $S$ & Baryon & Flavour structure ($I_3 = I$)		\\ \hline
$1/2$ & $0$ & $N$ &
$\frac{1}{\sqrt{2}} (\mid u d u \rangle - \mid d u u \rangle)$	\\
$3/2$ & $0$ & $\Delta$ & $\mid uuu \rangle$		\\
$0$ & $-1$ & $\Lambda$ &
$\frac{1}{\sqrt{2}} (\mid u d s \rangle - \mid d u s \rangle)$	\\
$1$ & $-1$ &  $\Sigma$ & $\mid u u s \rangle$			\\
$1/2$ & $-2$ & $\Xi$ & $\mid u s s \rangle$		\\
$0$ & $-3$ & $\Omega$ & $\mid sss \rangle$
\end{tabular}
\end{center}
\end{table}

The flavour structure is straightforward to implement in the
calculation of the correlation functions by only including the
appropriate contractions in the Wick expansion of
\eqn{eq:corrs_cons}.  

\subsection{Angular Momentum and Lattice QCD}
More delicate is the identification of the spin of particles in a
lattice calculation.  For the discussion of the spectrum, we are
principally interested in the study of states at rest, for which the
relevant continuum symmetry for classifying states is their properties
under rotations, described by the group $SU(2)$.  The irreducible
representations of $SU(2)$ are labeled by the total spin $J$, and the
projection of spin along a particular axis, say $J_3$, in the manner
of isospin.  We have already seen that we can impose exact isospin
symmetry in our calculations, but the use of a hypercubic lattice has
the consequence that we no longer have exact three-dimensional
rotational symmetry, but rather the symmetries of the
three-dimensional cubic group of the three-dimensional spatial
lattice, the octahedral group ${\cal O}$~\cite{elliott}.

In contrast to the continuum rotation group, ${\cal O}$ is a finite
group.  It contains a total of 48 elements, and has a total of five
single-valued irreducible representations (IR), corresponding to states of
integer spin, and only three double-valued IRs,
corresponding to states of half-integer spin.  Thus each lattice IR
is an admixture of different values of $J$, listed for the
double-valued representations in Table~\ref{tab:irs}~\cite{johnson82}.
\begin{table}
\begin{center}
\caption{Irreducible representations of the octahedral group,
	listing the number of elements and the spin content.}
\label{tab:irs}
\begin{tabular}{c|c|l}
Representation\ & Dimension\ & Spin ($J$)		\\ \hline
$G_1$ & 2 & $1/2, 7/2, 9/2, \dots$\\
$G_2$ & 2 & $5/2, 7/2, 11/2, \dots$\\
$H$ & 4 &   $3/2, 5/2, 7/2,\dots$
\end{tabular}
\end{center}
\end{table}

Thus in any lattice calculation at a fixed value of the lattice
spacing, we will aim to construct operators transforming irreducibly
according to one of the IRs of Table~\ref{tab:irs}, and extract the
spectrum of states within each of these IRs. We will only be able to
identify the angular momentum of the various states when we look for
commonality between the masses extracted from the various IRs in the
approach to the continuum limit.  Thus, for example, the continuum
state of spin $5/2$ has two degrees of freedom associated with $G_2$,
and the remaining four degrees of freedom associated with $H$.

So far, most lattice calculations have employed local, $S$-wave
propagators for the quarks, and as we shall see later, have employed
operators transforming according to the IRs $G_1$, for
spin $\frac{1}{2}$, or $H$, for spin $\frac{3}{2}$, and have
implicitly assumed that these states with these spins are indeed the
lightest states in their respective representations,
as observed in the physical hadron spectrum.   The
technology required to construct general baryon interpolating
operators transforming irreducibly under $O$ has now be
developed~\cite{morningstar03,lhpcops03}, and a preliminary attempt at
the spectrum using $P$-wave quark propagators has been made in
ref.~\cite{sato03}.

\subsection{Parity}

The remaining symmetry that we must consider is parity, corresponding
to the spatial-inversion operator $I_s$.  Clearly, this is a good
symmetry on our hypercubic lattice, yielding the \textit{point group}
$O_h$, with 96 elements, and an additional subscript $g$ or $u$ on our
irreducible representations corresponding to positive- and
negative-parity representations respectively.  One subtlety arises
when one considers the determination of the spectrum on a lattice with
either periodic or anti-periodic boundary conditions in the temporal
direction.  In that case, for identical source and sink we have 
\begin{eqnarray}
G(t) & = & \sum_{\vec{x}} \langle {\cal O}_{+}(\vec{x},t) \overline{\cal
  O}_{+}(0) \rangle\nonumber\\
 & \longrightarrow & \sum_{P_{+}} |A_{P_{+}}|^2 e^{- M_{P_{+}} t} \pm \sum_{P_{-}}
  | A_{P_{-}}|^2 e^{- M_{P_{-}} (N_t - t)}\ ,
\label{periodic}
\end{eqnarray}
where ${\cal O}_{+}$ is an interpolating operator designed to isolate
states of positive parity propagating in the forward time direction,
$N_t$ is the temporal extent of the lattice, $A_P$ is the amplitude of
a state $P$, and the subscripts $+~\mbox{and}~-$ denote contributions
of states of positive or negative parity respectively.  As discussed
further in Sec.~\ref{spin12LT}, a similar expression may be obtained
for the opposite parity states.  Thus we see that a complete
delineation of the states on a periodic lattice only occurs as $N_t
\to \infty$.  The superposition of results from periodic and
anti-periodic boundary conditions can be used to eliminate the second
term of Eq.~(\ref{periodic}).  Alternatively, a fixed boundary
condition can be used to eliminate the second term of
Eq.~(\ref{periodic}) by preventing states from crossing the temporal
boundary of the lattice.

\subsection{Smearing and Extended Interpolating Fields}

It is common to perform some smearing in the spatial
dimensions at the source to increase the overlap of the
interpolating operators with the ground states.
Here we describe one such technique to do this: gauge-invariant
Gaussian smearing \cite{Gusken:qx}.

The source-smearing technique \cite{Gusken:qx} starts with a point
source, 
\be
\psi_{0\, \alpha}^{\phantom{0}\, a}({\vec x}, t) = \delta^{ac}
\delta_{\alpha\gamma} \delta_{{\vec x},{\vec x}_0} \delta_{t,t_0}
\label{ptsource}
\ee
for source colour $c$, Dirac $\gamma$, position ${\vec x}_0$
and time $t_0$, and proceeds via the iterative scheme, 
\[
\psi_i({\vec x},t) = \sum_{{\vec x}'} F({\vec x},{\vec x}') \,
\psi_{i-1}({\vec x}',t) \, ,
\]
where
\[
F({\vec x},{\vec x}') = \frac{1}{(1+\alpha)} \left( \delta_{{\vec x},
    {\vec x}'} + 
  \frac{\alpha}{6} \sum_{\mu=1}^3 \left [ U_\mu({\vec x},t) \,
\delta_{{\vec x}',
{\vec x}+\widehat\mu} + 
U_\mu^\dagger({\vec x}-\widehat\mu,t) \, \delta_{{\vec x}', {\vec
    x}-\widehat\mu} \right ] \right) \, . 
\]
Repeating the procedure $N$ times gives the resulting fermion source,
\be
\psi_N({\vec x},t) = \sum_{{\vec x}'} F^N({\vec x},{\vec x}') \,
\psi_0({\vec x}',t) \, .
\ee
The parameters $N$ and $\alpha$ govern the size and shape of the
smearing function.
The propagator $S$ is obtained from the smeared source by
solving 
\be
M_{\alpha\beta}^{ab}\, S_{\beta\gamma}^{bc} =
\psi_{\alpha}^{a}\, ,
\ee
for each colour, Dirac source $c,\, \gamma$ respectively of
Eq.~(\ref{ptsource}) via a standard matrix inverter such as the
BiStabilised Conjugate Gradient algorithm \cite{BiCG}.

\section{Operators for Spin-${1\over 2}$ and Spin-${3\over 2}$ Baryons}

We will now see how the considerations described above apply to the
construction of operators for the spin-${1\over 2}$ and spin-${3 \over 2}$
baryons.
Initially we will construct the operators based on analogy with the
construction of the interpolating operators of the continuum, first
for spin ${1\over 2}$ and then for spin ${3 \over 2}$.
We will then briefly describe the construction of the operators that
transform irreducibly under the lattice symmetries directly, without
reference to the  continuum discussion.

\subsection{Spin-${1\over 2}$ Baryons: General Considerations}
\label{spin12LT}

Following standard notation, we define a two-point correlation
function for a spin-$\frac{1}{2}$ baryon $B$ as
\be
G_B(t,{\vec p})
\equiv \sum_{\vec{x}}\, e^{-i {\vec p} \cdot {\vec x}}
\left\langle \Omega \left| 
\chi_B(x)\bar\chi_B(0)
\right| \Omega \right\rangle
\label{2ptfunc}
\ee
where $\chi_B$ is a baryon interpolating field and where we have
suppressed Dirac indices.  All formalism for correlation functions and
interpolating fields presented in this article is carried out using
the Dirac representation of the Dirac $\gamma$-matrices.  The choice of
interpolating field $\chi_B$ is discussed in Section~\ref{spin12IF}
below.
The overlap of the interpolating field
$\chi_B$ with positive or negative parity states $| B^\pm \rangle$ is
parameterised by a coupling strength $\lambda_{B^\pm}$ which is
complex in general and which is defined by
%
\begin{eqnarray}
\left\langle \Omega \left|\, \chi_B(0)\, \right| B^+ ,p,s \right\rangle
\!\!&=&\!\! \lambda_{B^+} \sqrt{M_{B^+} \over E_{B^+}}\ u_{B^+}(p,s)\, ,         \\
\left\langle \Omega \left|\, \chi_B(0)\, \right| B^- ,p,s \right\rangle
\!\!&=&\!\! \lambda_{B^-} \sqrt{M_{B^-} \over E_{B^-}}\, \gamma_5
u_{B^-}(p,s)\, ,
\end{eqnarray}
%
where $M_{B^\pm}$ is the mass of the state $B^\pm$,
$E_{B^\pm} = \sqrt{M_{B^\pm}^2 + {\vec p\,}^2}$ is its energy,
and $u_{B^\pm}(p,s)$ is a Dirac spinor with normalisation
$\overline{u}_{B^\pm}^{\alpha}(p,s)u_{B^\pm}^{\beta}(p,s) =
\delta^{\alpha\beta}$.
For large Euclidean time, the correlation function can be written as a
sum of the lowest energy positive and negative parity contributions
%
\be
\label{G+-}
G_B(t,{\vec p})
\approx \lambda_{B^+}^2
        { \left( \gamma \cdot p + M_{B^+} \right) \over 2 E_{B^+} }
        e^{- E_{B^+} \, t}
 + \lambda_{B^-}^2
        { \left( \gamma \cdot p - M_{B^-} \right) \over 2 E_{B^-} }
        e^{- E_{B^-} \, t}\ ,
\label{cfunc}
\ee
%
when a fixed boundary condition in the time direction is used to 
remove backward propagating states.
The positive and negative parity states are isolated by taking the trace
of $G_B$ with the operator $\Gamma_+$ and $\Gamma_-$ respectively, where
\begin{eqnarray}
\Gamma_\pm &=&
{1\over 2} \left( 1 \pm {M_{B^\mp} \over E_{B^\mp}} \gamma_4 \right)\ .
\label{pProjOp}
\end{eqnarray}
For $\vec{p}=0$, the energy $E_{B^{\mp}} = M_{B^{\mp}}$, so that
$\Gamma_{\mp}^2 = \Gamma_{\mp}$ and the $\Gamma_{\mp}$ are parity
projectors In this case, positive parity states propagate in the (1,
1) and (2, 2) elements of the Dirac $\gamma$-matrix of
Eq.~(\ref{cfunc}), while negative parity states propagate in the (3,
3) and (4, 4) elements, and the masses of $B^{\mp}$ may be isolated.

%
%

\subsection{Spin-${1\over 2}$ Baryons: Interpolating Fields}
\label{spin12IF}

There are two types of interpolating fields which have
commonly been used in the literature.
The notation adopted here is similar to that of Ref.~\cite{LWD}.
To access the positive parity proton we use as interpolating fields
\begin{eqnarray}
\label{chi1p}
\chi_1^{p +}(x)
&=& \epsilon_{abc}
\left( u^T_a(x)\ C \gamma_5\ d_b(x) \right) u_c(x)\ ,
\end{eqnarray}
and
\begin{eqnarray}
\label{chi2p}
\chi_2^{p +}(x)
&=& \epsilon_{abc}
\left( u^T_a(x)\ C\ d_b(x) \right) \gamma_5\ u_c(x)\ ,
\end{eqnarray}
where the fields $u$, $d$ are evaluated at Euclidean space-time point
$x$, $C$ is the charge conjugation matrix, $a, b$ and $c$ are
colour labels, and the superscript $T$ denotes the transpose.
These interpolating fields transform as spinors under a parity
transformation.
That is, if the quark fields $q_a(x)\ (q=u,d, \cdots)$ transform as
\be
{\cal P} q_a(x) {\cal P}^\dagger = +\gamma_0 q_a(\tilde{x})\ ,
\label{paritytransform}
\ee
where $\tilde{x} = (x_0 , -\vec{x})$, then
$$
{\cal P} \chi^{p +}(x) {\cal P}^\dagger
 = +\gamma_0 \chi^{p +}(\tilde{x})\ .
$$

For convenience, we introduce the shorthand notation
%
\begin{eqnarray}
{\cal G}(S_{f_1},S_{f_2},S_{f_3}) &\equiv&
   \epsilon^{abc} \epsilon^{a'b'c'} \biggl\{
   S_{f_1}^{a a'}(x,0) \, {\rm tr} \left [ S_{f_2}^{b b' \, T}(x,0)
   S_{f_3}^{c c'}(x,0)
   \right ] \quad \nonumber\\ &+& S_{f_1}^{a a'}(x,0) \, S_{f_2}^{b b' \, T}(x,0) \,
   S_{f_3}^{c c'}(x,0) \biggr \} ,
\label{F}
\end{eqnarray}
where $S^{a a'}_{f_{1-3}}(x,0)$ are the quark propagators in the
background link-field configuration $U$ corresponding to flavours
$f_{1-3}$.
This allows us to express the correlation functions in a compact form.
The associated correlation function for $\chi_1^{p+}$ can be
written as
\begin{equation}
G^{p+}_{11}(t,\vec p; \Gamma) = \left\langle \sum_{\vec x}
  e^{-i \vec p \cdot \vec x} {\rm tr}
  \left[ -\Gamma \, \, {\cal G}
    \left( S_u, \, \widetilde C S_d {\widetilde C}^{-1}, \, S_u \right)
  \right] \right\rangle\ ,
\label{p11CF}
\end{equation}
%
where $\langle\cdots\rangle$ is the ensemble average over the link fields,
$\Gamma$ is the $\Gamma_{\pm}$ projection operator from
Eq.~(\ref{pProjOp}), and $\widetilde C = C\gamma_5$.
For ease of notation, we will drop the angled brackets,
$\langle\cdots\rangle$, and all the following correlation functions will
be understood to be ensemble averages.  
For the $\chi_2^{p+}$ interpolating field, one can similarly write
%
\begin{equation}
G^{p+}_{22}(t,\vec p; \Gamma) = \sum_{\vec x}
  e^{-i \vec p \cdot \vec x} {\rm tr}
  \left[ - \Gamma \, \, {\cal G}
    \left( \gamma_5 S_u \gamma_5, \, \widetilde C S_d {\widetilde C}^{-1},
           \, \gamma_5 S_u \gamma_5
    \right)
  \right]\ ,
\label{p22CF}
\end{equation}
while the interference terms from these two interpolating fields are
given by
\begin{eqnarray}
G^{p+}_{12}(t,\vec p; \Gamma) 
&=& \sum_{\vec x} e^{-i \vec p \cdot \vec x} \hfill  \hfill
   {\rm tr} \biggl [ - \Gamma \, \, \biggl \{  {\cal G} \left ( S_u \gamma_5,\,
   \widetilde C S_d {\widetilde C}^{-1}, \,  S_u \gamma_5\right )
   \biggr \} \biggr ] , 
\label{p12CF} \\
G^{p+}_{21}(t,\vec p; \Gamma)
&=& \sum_{\vec x} e^{-i \vec p \cdot \vec x} \hfill  \hfill
   {\rm tr} \biggl [ - \Gamma \, \, \biggl \{  
   {\cal G} \left ( \gamma_5 S_u,\,
   \widetilde C S_d {\widetilde C}^{-1}, \,  \gamma_5 S_u
     \right ) \biggr \} \biggr ] . 
\label{p21CF}
\end{eqnarray}

The neutron interpolating field is obtained via the exchange 
$u\leftrightarrow d$, and the
strangeness --2, $\Xi$ interpolating field by
replacing the doubly represented $u$ or $d$ quark fields in
Eqs.~(\ref{chi1p}) and (\ref{chi2p}) by $s$ quark fields. 
The $\Sigma$ and $\Xi$ interpolators are discussed in detail below.

As pointed out in Ref.~\cite{DEREK}, because of the Dirac structure
of the ``diquark'' in the parentheses in Eq.~(\ref{chi1p}), in the
Dirac representation the field
$\chi_1^{p+}$ involves both products of
{\em upper}~$\times$~{\it upper}~$\times$~{\it upper} and
{\it lower}~$\times$~{\it lower}~$\times$~{\it upper} 
components of spinors for positive parity baryons, so that in
the nonrelativistic limit $\chi_1^{p+} = {\cal O}(1)$.
Here upper and lower refer to the large and small spinor components in
the standard Dirac representation of the $\gamma$-matrices. 
Furthermore, since the ``diquark'' couples to total spin 0, one expects
an attractive force between the two quarks, and hence better overlap with a lower energy
state than with a state in which two quarks do not couple to spin 0.

The $\chi_2^{p+}$ interpolating field, on the other hand, is known to
have little overlap with the nucleon ground state \cite{DEREK,BOWLER}.
Inspection of the structure of the Dirac $\gamma$-matrices in
Eq.~(\ref{chi2p}) reveals that it involves only products of
{\it upper}~$\times$~{\it lower}~$\times$~{\it lower} components for
positive parity baryons,
so that $\chi_2^{p+} = {\cal O}(p^2 /E^2)$ vanishes in the
nonrelativistic limit.  As a result of the mixing of upper and lower
components, the ``diquark'' term contains a factor $\vec \sigma \cdot
\vec p$, meaning that the quarks no longer couple to spin 0, but are
in a relative $L=1$ state.  One expects therefore that two-point
correlation functions constructed from the interpolating field
$\chi_2^{p+}$ are dominated by larger mass states than those arising
from $\chi_1^{p+}$ at early Euclidean times.

While the masses of negative parity baryons are obtained directly 
from the (positive parity) interpolating fields in Eqs.~(\ref{chi1p}) and
(\ref{chi2p}) by using the parity projectors $\Gamma_\pm$, 
it is instructive nevertheless to examine the general properties
of the negative parity interpolating fields.
Interpolating fields with strong overlap with the negative parity
proton can be constructed by multiplying the previous positive parity
interpolating fields by
$\gamma_5$, $\chi_{1,2}^{p-} \equiv \gamma_5\ \chi_{1,2}^{p+}$.
In contrast to the positive parity case, both the interpolating fields
$\chi_1^{p -}$ and $\chi_2^{p -}$ mix upper and lower components, and
consequently both $\chi_1^{p -}$ and $\chi_2^{p -}$ are ${\cal O}(p/E)$.

Physically, two nearby $J^P = {1\over2}^-$ states are observed in the
excited nucleon spectrum.  In simple quark models, the splitting of
these two orthogonal states is largely attributed to the extent to
which the wave function is composed of scalar diquark configurations.
It is reasonable to expect $\chi_1^{p-}$ to have better overlap with
scalar diquark dominated states, and thus provide a lower effective
mass in the moderately large Euclidean time regime explored in lattice
simulations.  If the effective mass associated with the $\chi_2^{p-}$
correlator is larger, then this would be evidence of significant
overlap of $\chi_2^{p-}$ with the higher lying $N^{{1\over2}-}$
states.  In this event, a correlation matrix analysis (see Appendix)
can be used to isolate these two states \cite{FP,CSSM}.

Interpolating fields for the other members of the flavour SU(3) octet
are constructed along similar lines.  For the positive parity $\Sigma^0$
hyperon one uses \cite{LWD}

\begin{eqnarray}
\hspace*{-0.5cm}
\chi_1^{\Sigma}(x)
&=& {1\over\sqrt{2}} \epsilon_{abc}
\Big\{ \left( u^T_a(x)\ C \gamma_5\ s_b(x) \right) d_c(x)
     + \left( d^T_a(x)\ C \gamma_5\ s_b(x) \right) u_c(x)
\Big\}\, ,
\label{chi1S} \\
\hspace*{-0.5cm}
\chi_2^{\Sigma}(x)
&=& {1\over\sqrt{2}} \epsilon_{abc}
\Big\{ \left( u^T_a(x)\, C \, s_b(x) \right)\,\gamma_5 \,
  d_c(x)
     + \left( d^T_a(x)\, C \, s_b(x) \right)\,\gamma_5\, u_c(x)
\Big\}\, .
\label{chi2S}
\end{eqnarray}
Interpolating fields for accessing other charge states of
$\Sigma$ are obtained by $d\rightarrow u$ or $u\rightarrow d$,
producing correlation functions analogous to those in
Eqs.~(\ref{p11CF}) through (\ref{p21CF}).  Note that $\chi_1^{\Sigma}$
transforms as a triplet under SU(2) isospin.  An SU(2) singlet
interpolating field can be constructed by replacing $``+"
\longrightarrow ``-"$ in Eqs.~(\ref{chi1S}) and (\ref{chi2S}).
For the SU(3) octet $\Lambda$ interpolating field (denoted by
``$\Lambda^8$''), one has
%
\begin{eqnarray}
\chi_1^{\Lambda^8}(x)
&=& {1\over\sqrt{6}} \epsilon_{abc}
\left\{ 2
   \left( u^T_a(x)\ C \gamma_5\ d_b(x) \right) s_c(x)
+\ \left( u^T_a(x)\ C \gamma_5\ s_b(x) \right) d_c(x)\
\right.         \nonumber\\
& & \hspace*{3cm}
\left.
-\ \left( d^T_a(x)\ C \gamma_5\ s_b(x) \right) u_c(x)
\right\}\ ,
\label{chi1l8} \\
\chi_2^{\Lambda^8}(x)
&=& {1\over\sqrt{6}} \epsilon_{abc}
\Big\{2
   \left( u^T_a(x)\, C \, d_b(x) \right)\,\gamma_5 \, s_c(x)
+ \left( u^T_a(x)\, C \, s_b(x) \right)\,\gamma_5 \, d_c(x)\
         \nonumber\\
& & \hspace*{3cm}
-\ \left( d^T_a(x)\, C \, s_b(x) \right)\,\gamma_5 \, u_c(x)
\Big\}\, ,
\end{eqnarray}
which leads to the correlation function
\begin{eqnarray}
\lefteqn{ G^{\Lambda^8}_{11}(t,\vec p; \Gamma)  = 
  {1 \over 6} \sum_{\vec x} e^{-i \vec p \cdot \vec x}} \nonumber \\
   & \times 
   {\rm tr} \biggl [  -\Gamma
\!\!  \biggl \{&  \!
     2 \,  {\cal G} \left ( S_s, \, S_d, \, \widetilde C S_u
   \widetilde C^{-1} \right )
   + 2 \, {\cal G} \left ( S_s, \, S_u, \, \widetilde C S_d
    \widetilde C^{-1} \right ) \nonumber\\
\!\! & &+ \! 2 \, {\cal G} \left ( S_d, \, S_s, \, \widetilde C S_u
    \widetilde C^{-1}\right )
   + 2 \, {\cal G} \left ( S_u, \, S_s, \, \widetilde C S_d
    \widetilde C^{-1}\right ) \nonumber\\
\!\! & &- \! \phantom{2} \, {\cal G} \left ( S_d, \, S_u, \, \widetilde C S_s
    \widetilde C^{-1}\right )
   - \phantom{2} \, {\cal G} \left ( S_u, \, S_d, \, \widetilde C S_s
    \widetilde C^{-1} \right )
   \biggr \} \biggr ]\ ,
\end{eqnarray}
%
and similarly for the correlation functions $G^{\Lambda^8}_{22}$,
$G^{\Lambda^8}_{12}$ and $G^{\Lambda^8}_{21}$.

The interpolating fields for the SU(3) flavour singlet (denoted by
``$\Lambda^1$'') are given by \cite{LWD}
\begin{eqnarray}
\chi_1^{\Lambda^1}(x)
&=& -2\ \epsilon_{abc}
\left\{ -
   \left( u^T_a(x)\ C \gamma_5\ d_b(x) \right) s_c(x)
+\ \left( u^T_a(x)\ C \gamma_5\ s_b(x) \right) d_c(x)
\right.                 \nonumber\\
& & \hspace*{3cm}
\left.
-\ \left( d^T_a(x)\ C \gamma_5\ s_b(x) \right) u_c(x)
\right\}\ ,
\label{chi1l1} \\
\chi_2^{\Lambda^1}(x)
&=& -2\ \epsilon_{abc}
\Big\{ -
   \left( u^T_a(x)\, C \, d_b(x) \right)\,\gamma_5 \, s_c(x) 
+\left( u^T_a(x)\, C \, s_b(x) \right)\,\gamma_5 \, d_c(x)
                 \nonumber\\
& & \hspace*{3cm}
- \left( d^T_a(x)\, C \, s_b(x) \right)\,\gamma_5 \, u_c(x)
\Big\}\ ,
\end{eqnarray}
where the last two terms are common to both $\chi_1^{\Lambda^8}$ and
$\chi_1^{\Lambda^1}$.
The correlation function resulting from this field involves quite a few
terms,
%
\begin{eqnarray}
G^{\Lambda^{1}}_{11}(t,\vec p; \Gamma) =\nonumber\\
   \sum_{\vec x} e^{-i \vec p \cdot \vec x} 
   {\rm tr} \biggl [ -\Gamma \, \, \biggl \{ & 
   \gamma_5 S_s^{a a'} \, \widetilde C S_d^{c c' \, T} \widetilde C^{-1}
        \,  S_u^{b b'} \gamma_5
+  \gamma_5 S_u^{a a'} \, \widetilde C S_d^{c c' \, T} \widetilde C^{-1}
        \,  S_s^{b b'} \gamma_5 \cr
&\!\! +\  \gamma_5 S_s^{a a'} \, \widetilde C S_u^{c c' \, T} \widetilde C^{-1}
        \,  S_d^{b b'} \gamma_5
+  \gamma_5 S_d^{a a'} \, \widetilde C S_u^{c c' \, T} \widetilde C^{-1}
        \,  S_s^{b b'} \gamma_5 \cr
&\!\! +\  \gamma_5 S_u^{a a'} \, \widetilde C S_s^{c c' \, T} \widetilde C^{-1}
        \,  S_d^{b b'} \gamma_5
+  \gamma_5 S_d^{a a'} \, \widetilde C S_s^{c c' \, T} \widetilde C^{-1}
        \,  S_u^{b b'} \gamma_5 \cr
& -\  \gamma_5 S_s^{a a'} \gamma_5 \  {\rm tr} \left [
   S_d^{b b'} \, \widetilde C S_u^{c c' \, T} \widetilde C^{-1} \right ] \cr
& -\  \gamma_5 S_u^{a a'} \gamma_5 \  {\rm tr} \left [
   S_s^{b b'} \, \widetilde C S_d^{c c' \, T} \widetilde C^{-1} \right ] \cr
& -\  \gamma_5 S_d^{a a'} \gamma_5 \  {\rm tr} \left [
   S_u^{b b'} \, \widetilde C S_s^{c c' \, T} \widetilde C^{-1} \right ]
   \biggr \} \biggr ] .
\end{eqnarray}
%
In order to test the extent to which SU(3) flavour symmetry is valid in
the baryon spectrum, one can construct another $\Lambda$ interpolating
field composed of the terms common to $\Lambda^1$ and $\Lambda^8$,
which does not make any assumptions about the SU(3) flavour symmetry
properties of $\Lambda$.
We define
\begin{eqnarray}
\hspace*{-0.5cm}
\chi_1^{\Lambda^c}(x)
&=& {1\over\sqrt{2}} \epsilon_{abc}
\Big\{
   \left( u^T_a(x)\ C \gamma_5\ s_b(x) \right) d_c(x)
- \left( d^T_a(x)\ C \gamma_5\ s_b(x) \right) u_c(x)
\Big\}\, , 
\label{chi1lc}\\
\hspace*{-0.5cm}
\chi_2^{\Lambda^c}(x)
&=& {1\over\sqrt{2}} \epsilon_{abc}
\Big\{
   \left( u^T_a(x)\, C \, s_b(x) \right)\,\gamma_5 \, d_c(x)
- \left( d^T_a(x)\, C \, s_b(x) \right)\,\gamma_5 \, u_c(x)
\Big\}\, ,
\label{chi2lc}
\end{eqnarray}
to be our ``common'' interpolating fields which are the isosinglet
analogue of $\chi_1^\Sigma$ and $\chi_2^\Sigma$ in Eqs.~(\ref{chi1S})
and (\ref{chi2S}).
Such interpolating fields may be useful in determining the nature of the
$\Lambda^*(1405)$ resonance, as they allow for mixing between singlet and
octet states induced by SU(3) flavour symmetry breaking.
To appreciate the structure of the ``common'' correlation function, one
can introduce the function
%
\begin{eqnarray}
 \overline {\cal G}(S_{f_1},S_{f_2},S_{f_3}) =
   \epsilon^{abc} \epsilon^{a'b'c'} &\biggl \{ &
   S_{f_1}^{a a'}(x,0) \, {\rm tr} \left [ S_{f_2}^{b b' \, T}(x,0)
   S_{f_3}^{c c'}(x,0) \right ] \nonumber \\
 &-& S_{f_1}^{a a'}(x,0) \, S_{f_2}^{b b' \, T}(x,0) \,
   S_{f_3}^{c c'}(x,0) \biggr \} ,
\end{eqnarray}
which is recognised as ${\cal G}$ in Eq.~(\ref{F}) with the relative
sign of the two terms changed.
With this notation, the correlation function corresponding to the
$\chi_1^{\Lambda^c}$ interpolating field is
\begin{eqnarray}
 G_{11}^{\Lambda_{C}}(t,\vec p; \Gamma) &=&
   {1 \over 2} \sum_{\vec x} e^{-i \vec p \cdot \vec x}
   {\rm tr} \Biggl [ -\Gamma \, \, \Bigl \{ \overline {\cal G} \left ( S_d,
   \, \widetilde C S_s \widetilde C^{-1} ,\, S_u \right ) 
   \nonumber \\
&& \qquad\qquad + \overline {\cal G} \left ( S_u, \,
   \widetilde C S_s \widetilde C^{-1} ,\,
   S_d \right ) \Bigr \} \Biggr ]\ ,
\end{eqnarray}
and similarly for the correlation functions involving the
$\chi_2^{\Lambda^c}$ interpolating field.

\subsection{Spin-${3\over 2}$ Baryons}
\label{spin32sec}


In this section we extend the discussion from the previous section to
the spin-$\frac{3}{2}$ baryon sector. The
mass of a spin-${3\over 2}$ baryon on the lattice is obtained from
the two-point correlation function $G_{\mu\nu}$ \cite{CSSM32,LDW},
\be
G_{\mu\nu}(t, \vec p; \Gamma) = {\rm tr} \left[ \Gamma {\cal
    G}_{\mu\nu}(t, \vec p) \right]\, ,
\label{2ptfunc32}
\ee
where
\begin{equation}
{\cal G}_{\mu\nu}^{\alpha\beta}(t, \vec p)
= \sum_{\vec x} e^{-i \vec p \cdot \vec x }\ 
  \langle \Omega |\ T \left( \chi_\mu^\alpha (x)\ \overline \chi_\nu^\beta
    (0)  \right)
  | \Omega \rangle\ ,
\label{CFunc}
\end{equation}
and $\chi_\mu^\alpha$ is a spin-${3\over 2}$ interpolating field,
$\Gamma$ is a matrix in Dirac space with $\alpha, \beta$ Dirac indices,
and $\mu, \nu$ Lorentz indices.

The interpolating field operator used in the literature for accessing the
isospin-${1\over 2}$, spin-${3\over 2}$, positive parity (charge $+1$)
state is \cite{CSSM32,LeeSpin32,IOFFE} 
\be
\chi^p_{\mu} = \epsilon^{abc} 
\left( u^{Ta}(x)\ C \gamma_5 \gamma^\nu\ d^b(x) \right)
\left( g_{\mu\nu} - {1 \over 4} \gamma_\mu \gamma_\nu \right)
\gamma_5 u^c(x)\ .
\label{N32IFfull}
\ee
As pointed out in Section~\ref{spin12LT}, all discussions of interpolating
fields are carried out using the Dirac representation of the
$\gamma$-matrices.
This exact isospin-${1\over 2}$ interpolating field has overlap with
both spin-${3\over 2}$ and spin-${1\over 2}$ states and with states of
both parities. The resulting correlation function will thus require
both spin and parity projection.
The charge neutral interpolating field is obtained by interchanging
$u \leftrightarrow d$.
This interpolating field transforms as a Rarita-Schwinger operator under
parity transformations. That is, if the quark field operators transform as
in Eq.~(\ref{paritytransform}), then
$$
{\cal P} \chi^{N}_\mu(x) {\cal P}^\dagger
 = +\gamma_0 \chi^{N}_\mu(\tilde{x})\ ,
$$
and similarly for the Rarita-Schwinger operator
\be
{\cal P} u_\mu (x) {\cal P}^\dagger = +\gamma_0 u_\mu (\tilde{x})\ ,
\ee
which will be discussed later.

The computational cost of evaluating each of the Lorentz combinations in
Eq.~(\ref{N32IFfull}) is relatively high --- about 100 times that for the
ground state nucleon \cite{LL}.
Consequently, in order to maximise statistics it is common to consider
only the leading term proportional to $g_{\mu\nu}$ in the
interpolating field,
\be
\chi^p_{\mu} \longrightarrow \epsilon^{abc}
\left( u^{Ta}(x)\ C \gamma_5 \gamma_\mu\ d^b(x) \right) \gamma_5 u^c(x)\ .
\label{N32IF}
\ee
This is sufficient since we will in either case need to perform a
spin-$\frac{3}{2}$ projection.

In order to show that the interpolating field defined in
Eq.~(\ref{N32IF}) has isospin $\frac{1}{2}$, we first consider the
standard proton interpolating field given in Eq.~(\ref{chi1p}) which we
know to have isospin $\frac{1}{2}$. Applying the isospin raising operator,
$I^+$, on $\chi^p$, one finds
\begin{eqnarray*}
I^+ \chi^p &=& \epsilon^{abc}(u^{Ta} C\gamma_5 u^b ) u^c \\
&=& \epsilon^{abc}(u^{Ta} C\gamma_5 u^b )^T u^c \\
&=& -\epsilon^{abc}(u^{Ta} C\gamma_5 u^b ) u^c \\
&=& 0\, .
\end{eqnarray*}
Similarly, for the interpolating field defined in Eq.~(\ref{N32IF}),
one has
\begin{eqnarray*}
I^+ \chi_{\mu}^p &=& \epsilon^{abc}(u^{Ta} C\gamma_5 \gamma_{\mu} u^b )
\gamma_5 u^c \\
&=& \epsilon^{abc}(u^{Ta} C\gamma_5 \gamma_{\mu} u^b )^T \gamma_5 u^c \\
&=& -\epsilon^{abc}(u^{Ta} C\gamma_5 \gamma_{\mu} u^b ) \gamma_5 u^c \\
&=& 0\, ,
\end{eqnarray*}
where we have used the representation-independent identities 
$C\gamma_{\mu}C^{-1} = -\gamma_{\mu}^T$, $C\gamma_5
C^{-1} = \gamma_5^T$, and the identities which hold in the Dirac
representation: $C^T = C^\dagger = C^{-1} = -C$ with $C =
i\gamma_2\gamma_0$ and $\gamma_5^T = \gamma_5$.

We note that $\bar{\chi}_\mu^p$, corresponding to $\chi_\mu^p$ in
Eq.~(\ref{N32IF}), is
\be
\bar{\chi}_\mu^p = \epsilon^{abc} \bar{u}^a \gamma_{ 5}(\bar{d}^b
\gamma_{\mu}\gamma_{ 5} C \bar{u}^{cT} )\,\ ,
\ee
so that
\begin{eqnarray}
\chi_{\mu}^p \bar{\chi}_{\nu}^p &
 = & \epsilon^{abc} \epsilon^{a'b'c'}
( u^{Ta}_{\alpha}[C\gamma_{5}\gamma_{\mu}]_{\alpha\beta} d^b_{\beta})
\gamma_{5} u^c_{\gamma} \bar{u}_{\gamma'}^{c'} \gamma_{5}
(\bar{d}^{b'}_{\beta'}[\gamma_{\nu}\gamma_{5} C]_{\beta' \alpha'}
\bar{u}^{T a'}_{\alpha'}) 		\nonumber \\
%
& \to &
\gamma_5 S_u \gamma_5 {\rm tr} \left[\gamma_5 S_u \gamma_5
\left(C\gamma_{\mu} S_d \gamma_{\nu} C\right)^T \right] + \nonumber \\
 & & \,\, \gamma_5 S_u \gamma_5
\left(C\gamma_{\mu} S_d \gamma_{\nu} C\right)^T \gamma_5 S_u \gamma_5\ ,
\end{eqnarray}
where the last line is the result after performing the Grassman
integration over the quark fields with the quark fields being replaced
by all possible pairwise contractions.

In deriving the $\Delta$ interpolating fields, it is simplest to begin
with the state containing only valence $u$ quarks, namely the
$\Delta^{++}$.
The interpolating field for the $\Delta^{++}$ resonance
is given by \cite{IOFFE}
\begin{equation}
\chi_\mu^{\Delta^{++}}(x)
= \epsilon^{abc}
\left( u^{Ta}(x)\ C \gamma_\mu\ u^b(x) \right) u^c(x)\ ,
\label{deltaIF}
\end{equation}
which transforms as a pseudovector under parity.
The interpolating field for a $\Delta^+$ state can be similarly
constructed \cite{LDW},
\be
\chi_\mu^{\Delta^{+}}(x)
= {1 \over \sqrt{3} } \; \epsilon^{abc}
\left[ 2 \left( u^{Ta}(x)\, C \gamma_\mu\, d^b(x) \right)\, u^c(x)
      + \left( u^{Ta}(x)\, C \gamma_\mu\, u^b(x) \right)\, d^c(x)
\right]\, .
\ee
Interpolating fields for other decuplet baryons are obtained by
appropriate substitutions of $u,\ d\ \to\ u,\ d$ or $s$
fields.

To project a pure spin-${3\over 2}$ state from the correlation function
$G_{\mu\nu}$, one needs to use an appropriate spin-${3\over 2}$ projection
operator \cite{BDM},
\begin{equation}
P^{3/2}_{\mu \nu}(p)
= g_{\mu \nu}
- {1 \over 3} \gamma_\mu \gamma_\nu
- {1 \over 3 p^2}
   \left( \gamma \cdot p\, \gamma_\mu\, p_\nu
        + p_\mu\, \gamma_\nu\, \gamma \cdot p
   \right)\ .
\end{equation}
The corresponding spin-${1\over 2}$ state can be projected by applying the
projection operator
\begin{equation}
P_{\mu\nu}^{1/2} = g_{\mu\nu} - P_{\mu\nu}^{3/2}\ .
\end{equation}
To use this operator and retain all Lorentz components,
one must calculate the full $4\times 4$ matrix in Dirac and Lorentz space.
However, to extract a mass, only one pair of Lorentz indices is needed, reducing
the amount of calculations required by a factor of four. The results
from Ref.~\cite{CSSM32} which are summarised in
Section~\ref{32results} are calculated from the third
row of the Lorentz matrix and using the projection
\be
G^s_{33} = \sum_{\mu,\nu = 1}^4 G_{3\mu}\, g^{\mu\nu}\, P_{\nu 3}^s\ ,
\label{SpinPrCF}
\ee
to extract the desired spin states, $s=\frac{1}{2}$ or $\frac{3}{2}$.
Following spin projection, the resulting correlation function, $G^s_{33}$,
still contains positive and negative parity states.


The interpolating field operators defined in Eqs.~(\ref{N32IFfull})
and (\ref{N32IF}) have overlap with both spin-$\frac{3}{2}$ and
spin-$\frac{1}{2}$ states with positive and negative parity.
The field $\chi_\mu$ transforms as a pseudovector under parity, as does
the Rarita-Schwinger spinor, $u_\mu$. Thus the overlap of $\chi_\mu$
with baryons can be expressed as
%
\begin{eqnarray}
\langle \Omega | \chi_{\mu} | N^{\frac{3}{2}+}(p,s) \rangle &=&
\lambda_{3/2^+} \sqrt{ {M_{3/2^+} \over E_{3/2^+}} }\ u_{\mu}(p,s)\ , \\
\langle \Omega | \chi_{\mu} | N^{\frac{3}{2}-}(p,s) \rangle &=&
\lambda_{3/2^-} \sqrt{ {M_{3/2^-} \over E_{3/2^-}} }\ \gamma_5
u_{\mu}(p,s)\ , \\
\langle \Omega | \chi_{\mu} | N^{\frac{1}{2}+}(p,s) \rangle &=&
(\alpha_{1/2^+}p_{\mu} + \beta_{1/2^+}\gamma_{\mu})
\sqrt{ {M_{1/2^+} \over E_{1/2^+}} }\ \gamma_5 u(p,s)\ , \label{third}\\
\langle \Omega | \chi_{\mu} | N^{\frac{1}{2}-}(p,s) \rangle &=&
(\alpha_{1/2^-}p_{\mu} + \beta_{1/2^-}\gamma_{\mu})
\sqrt{ {M_{1/2^-} \over E_{1/2^-}} }\ u(p,s)\ , \label{fourth}
\end{eqnarray}
%
where the factors $\lambda_B ,\, \alpha_B ,\,
\beta_B$ denote the coupling strengths of the interpolating field
$\chi_\mu$ to the baryon $B$.
For the expressions in Eqs.~(\ref{third}) and (\ref{fourth}),
we note that the 
spatial components of momentum, $p_i$, transform as vectors under
parity and commute with $\gamma_0$, whereas the $\gamma_i$ do not
change sign under parity but anti-commute with $\gamma_0$. Hence the
right-hand-side of Eq.~(\ref{third}) also transforms as a
pseudovector under parity in accord with $\chi_\mu$.

Similar expressions can also be written for $\bar{\chi}_\mu$,
%
%
\begin{eqnarray}
\langle N^{\frac{3}{2}+}(p,s) | \bar{\chi}_{\mu} | \Omega \rangle &=&
\lambda_{3/2^+}^* \sqrt{ {M_{3/2^+} \over E_{3/2^+}} }\
\bar{u}_{\mu}(p,s)\ ,\\
\langle N^{\frac{3}{2}-}(p,s) | \bar{\chi}_{\mu} | \Omega \rangle 
&=& -\lambda_{3/2^-}^* \sqrt{ {M_{3/2^-} \over E_{3/2^-}} }\
\bar{u}_{\mu}(p,s) \gamma_5\ ,\\
\langle N^{\frac{1}{2}+}(p,s) | \bar{\chi}_{\mu} | \Omega \rangle 
%
&=& - \sqrt{ {M_{1/2^+} \over E_{1/2^+}} }\ \bar{u}(p,s) \gamma_5
(\alpha^*_{1/2^+}p_{\mu} + \beta^*_{1/2^+}
\gamma_{\mu})\ , \\
\langle N^{\frac{1}{2}-}(p,s) | \bar{\chi}_{\mu} | \Omega \rangle 
%
&=& \sqrt{ {M_{1/2^-} \over E_{1/2^-}} }\ \bar{u}(p,s)
(\alpha^*_{1/2^-}p_{\mu} + \beta^*_{1/2^-}
\gamma_{\mu})\ .
\end{eqnarray}
%
Note that we are assuming identical sinks and sources in these
equations.  However, calculations often use a smeared source and a point
sink in which case $\lambda^* ,\ \alpha^*$ and $\beta^*$ are no longer
complex conjugates of $\lambda,\ \alpha$ and $\beta$ but are instead
replaced by $\overline{\lambda},\ \overline{\alpha}$ and
$\overline{\beta}$. 

We are now in a position to find the form of Eq.~(\ref{CFunc}) after we
insert a complete set of intermediate states $\left\{| B (p,s)
  \rangle\right\}$. 
The contribution to Eq.~(\ref{CFunc}) from each intermediate state
considered is given by 
\begin{eqnarray*}
\langle \Omega | \!\! 
&\chi_{\mu}& \!\! | N^{\frac{3}{2}+}(p,s) \rangle 
\langle N^{\frac{3}{2}+}(p,s) | \bar{\chi}_{\nu} | \Omega \rangle \\
\!\!\!\!\!&=&\!\!\! + \lambda_{3/2^+}\overline{\lambda}_{3/2^+}\,{M_{3/2^+}
  \over E_{3/2^+}} u_\mu (p,s) \bar{u}_\nu (p,s) \\
\!\!\!\!\!&=&\!\!\! - \lambda_{3/2^+}\overline{\lambda}_{3/2^+}\, {M_{3/2^+}
  \over E_{3/2^+}}  
{ (\gamma \cdot p + M_{3/2^+}) \over 2 M_{3/2^+} }
 \left\{ g_{\mu\nu}
        - { 1 \over 3} \gamma_\mu \gamma_\nu
        - { 2 p_\mu p_\nu \over 3 M^2_{3/2^+} }
        + { p_\mu \gamma_\nu - p_\nu \gamma_\mu \over 3 M_{3/2^+}}
  \right\} , \\
\langle \Omega | \!\! 
&\chi_{\mu}& \!\! | N^{\frac{3}{2}-}(p,s) \rangle 
\langle N^{\frac{3}{2}-}(p,s) | \bar{\chi}_{\nu} | \Omega \rangle \\
\!\!\!\!\!&=& \!\!\!
- \lambda_{3/2^-}\overline{\lambda}_{3/2^-}\, {M_{3/2^-} \over E_{3/2^-}} \gamma_5 u_\mu (p,s) \bar{u}_\nu (p,s)
\gamma_5 \\
%
\!\!\!\!\!&=&\!\!\! -\lambda_{3/2^-}\overline{\lambda}_{3/2^-}\, {M_{3/2^-}
  \over E_{3/2^-}}  
{ (\gamma \cdot p - M_{3/2^-}) \over 2 M_{3/2^-} }
  \left\{ g_{\mu\nu}
        - { 1 \over 3} \gamma_\mu \gamma_\nu
        - { 2 p_\mu p_\nu \over 3 M^2_{3/2^-} }
        - { p_\mu \gamma_\nu - p_\nu \gamma_\mu \over 3 M_{3/2^-}}
  \right\} , \\
\langle \Omega | \!\! 
&\chi_{\mu}& \!\! | N^{\frac{1}{2}+}(p,s) \rangle 
\langle N^{\frac{1}{2}+}(p,s) | \bar{\chi}_{\nu} | \Omega \rangle \\
\!\!\!\!\!&=& \!\!\!-{M_{1/2^+} \over E_{1/2^+}}
(\alpha_{1/2^+}p_{\mu} + \beta_{1/2^+}
\gamma_{\mu}) \gamma_5 \frac{\gamma\cdot p + M_{1/2^+}}{2M_{1/2^+}} \gamma_5
(\overline\alpha_{1/2^+}p_{\nu} + \overline\beta_{1/2^+}
\gamma_{\nu})\ , \\
\langle \Omega | \!\! 
&\chi_{\mu}& \!\! | N^{\frac{1}{2}-}(p,s) \rangle 
\langle N^{\frac{1}{2}-}(p,s) | \bar{\chi}_{\nu} | \Omega \rangle \\
\!\!\!\!\!&=& \!\!\!{M_{1/2^-} \over E_{-/2^+}}
(\alpha_{1/2^-}p_{\mu} + \beta_{1/2^-}
\gamma_{\mu}) \frac{\gamma\cdot p + M_{1/2^-}}{2M_{1/2^-}} 
(\overline\alpha_{1/2^-}p_{\nu} + \overline\beta_{1/2^-}
\gamma_{\nu})\ .
\end{eqnarray*}
To reduce computational expense, we consider the specific case when
$\mu = \nu = 3$ and in order to extract masses we require $\vec{p} =
(0,0,0)$. In this case we have the simple expressions
%
\begin{eqnarray}
\langle \Omega | \chi_{3} | N^{\frac{3}{2}+}(p,s) \rangle 
\langle N^{\frac{3}{2}+}(p,s) | \bar{\chi}_{3} | \Omega \rangle
&=& 
\lambda_{3/2^+}\overline{\lambda}_{3/2^+} \frac{2}{3}
\left( { \gamma_0 M_{3/2^+} + M_{3/2^+} \over 2 M_{3/2^+} } \right)\, , \\
\langle \Omega | \chi_{3} | N^{\frac{3}{2}-}(p,s) \rangle 
\langle N^{\frac{3}{2}-}(p,s) | \bar{\chi}_{3} | \Omega \rangle
&=& 
\lambda_{3/2^-}\overline{\lambda}_{3/2^-} \frac{2}{3}
\left( { \gamma_0 M_{3/2^-} - M_{3/2^-} \over 2 M_{3/2^-} } \right)\, , \\
\langle \Omega | \chi_{3} | N^{\frac{1}{2}+}(p,s) \rangle 
\langle N^{\frac{1}{2}+}(p,s) | \bar{\chi}_{3} | \Omega \rangle 
&=& - \beta_{1/2^+}\overline{\beta}_{1/2^+} \gamma_3 \gamma_5 \frac{\gamma_0 M_{1/2^+} +
  M_{1/2^+}}{2M_{1/2^+}} \gamma_5 \gamma_3 \nonumber \\
&=& + \beta_{1/2^+}\overline{\beta}_{1/2^+} \frac{\gamma_0 M_{1/2^+} +
  M_{1/2^+}}{2M_{1/2^+}}\, , \\
\langle \Omega | \chi_{3} | N^{\frac{1}{2}-}(p,s) \rangle 
\langle N^{\frac{1}{2}-}(p,s) | \bar{\chi}_{3} | \Omega \rangle 
&=& \beta_{1/2^-}\overline{\beta}_{1/2^-} \gamma_3 \frac{\gamma_0 M_{1/2^-} +
  M_{1/2^-}}{2M_{1/2^-}} \gamma_3 \nonumber \\
&=& + \beta_{1/2^-}\overline{\beta}_{1/2^-} \frac{\gamma_0 M_{1/2^-} -
  M_{1/2^-}}{2M_{1/2^-}}\, .
\end{eqnarray}
%
Therefore, in an analogous procedure to that used in
Section.~\ref{spin12LT}, where a fixed boundary condition is used in the
time direction, positive and negative parity states are obtained by
taking the trace of the spin-projected correlation function,
$G^s_{33}$, in Eq.~(\ref{SpinPrCF}) with the operator 
$\Gamma = \Gamma_{\pm}$, 
\be
G^{s\pm}_{33} = {\rm tr}\left[ \Gamma^{\pm}\, G^s_{33}\right]\, ,
\ee
where in this case (cf. Eq.~(\ref{pProjOp}))
\be
\Gamma_{\pm} = {1\over 2}\left( 1\pm \gamma_4 \right) .
\ee
The positive parity states propagate in the (1,1) and (2,2) elements of
the Dirac $\gamma$-matrix, while negative parity states propagate
in the (3,3) and (4,4) elements for both spin-$\frac{1}{2}$ and
spin-$\frac{3}{2}$ projected states. A similar treatment can be
carried out for the $\Delta$ interpolating fields but is left as an
exercise for the interested reader.

\subsection{Octahedral Group Irreducible Representations}

The procedure outlined above largely follows the continuum
construction of interpolating operators used in, say, QCD sum-rule
calculations.  An alternative approach is to construct operators
that lie in the irreducible representations (IRs) of the cubic
group $O_h$ of the lattice directly \cite{morningstar03,lhpcops03}.
To illustrate the latter method, we consider the construction
of the $I = 1/2, I_3 = 1/2$ nucleon interpolating operator,
following the discussion of Ref.~\cite{lhpcops03}.

The starting point is the formation of a set of elemental baryon operators
that are gauge invariant, and have the correct isospin, or flavour,
properties.  In the case of point-like quark fields, these are given by
\begin{equation}
\Phi_{\alpha\beta\gamma} = \epsilon^{abc}(u^a_{\alpha} d^b_{\gamma}
u^c_{\beta} - d^a_{\alpha} u^b_{\gamma} u^c_{\beta})\ ,\label{eq:phi}
\end{equation}
where the coordinate of the quark fields are suppressed, and $\alpha,
\beta, \gamma$ are spinor indices.  Examination of \eqn{eq:phi}
reveals the constraints $\Phi_{\alpha\beta\gamma} +
\Phi_{\gamma\beta\alpha} = 0$ and $\Phi_{\alpha\beta\gamma} +
\Phi_{\beta\gamma\alpha} + \Phi_{\gamma\alpha\beta} = 0$ so that there
are only 20 independent operators.

The central result needed to find operators transforming irreducibly
under the IRs of Table~\ref{tab:irs} is the projection formula
\begin{equation}
  B_i^{\Lambda\lambda F}\!(\vec{x})\!
 = \!\frac{d_\Lambda}{g_{O_h}}\!\!\!\sum_{R\in O_h}\!\!
  \!\!D^{(\Lambda)\ast}_{\lambda\lambda}(R)
   U_R B^F_i\!(\vec{x}) U_R^\dagger\ ,
\label{eq:project}\end{equation}
where $\Lambda$ refers to an $O_h$ IR, $\lambda$ is the IR row,
$g_{O_h}$ is the number of elements in $O_h$, $d_\Lambda$ is the
dimension of the $\Lambda$ IR, $D^{(\Lambda)}_{mn}(R)$ is a
$\Lambda$ representation matrix corresponding to group element $R$,
and $U_R$ is the quantum operator which implements the symmetry
operations; the temporal argument is suppressed.  Application of the
formula requires explicit representation matrices for each group
element.  These are applied to the 20 linearly independent operators
above, from which 20 linearly independent operators that transform
irreducibly under $O_h$ are identified.  The procedure is Dirac-basis
dependent, and linearly independent operators are shown in
Table~\ref{tab:ir_results}, using the DeGrand-Rossi representation
of the Dirac $\gamma$-matrices~\cite{lhpcops03}.

There are three embeddings of $G_1$ and a single embedding of $H$ for
both even parity ($g$) and odd parity ($u$); the two helicities of spin
$1/2$ correspond to the two rows of the two-dimensional representation
$G_{1g(1u)}$ while the four helicities of spin $3/2$ correspond to the
four rows of the four-dimensional representation $H$.  The three
embeddings of $G_1$ correspond to some linear combinations of the
operators of Eqs.~(\ref{chi1p})--(\ref{chi2p}) and the spin-$1/2$
projection of Eq.~(\ref{N32IFfull}); the embedding of $H$ corresponds to
the spin-$3/2$ projection.

Whilst the results in Table~\ref{tab:ir_results} may be less intuitive than
their counterparts derived earlier, the procedure can easily extended
to more complicated operators, including those with excited glue, or to
pentaquark operators.  Furthermore, we have seen explicitly that there
are three independent spin-$1/2$ operators.

\begin{table}
\begin{center}
\caption{Combinations of the operators $\Phi_{\alpha\beta\gamma}$
 in \protect\eqn{eq:project} 
 which transform irreducibly under $O_h$ for the DeGrand-Rossi
 representation of the Dirac $\gamma$-matrices, employed by LHPC and MILC
 Collaborations (see Ref.~\protect\cite{lhpcops03}).
\label{tab:ir_results}}
\renewcommand{\tabcolsep}{4mm} 
\begin{tabular}{ccc}\hline
 Irrep & Row &  Operators
    \\ \hline 
  $G_{1g}$  & 1 &  $\Phi_{112} + \Phi_{334}$  \\
  $G_{1g}$  & 2 &  $-\Phi_{221} - \Phi_{443}$  \\ \hline
  $G_{1g}$  & 1 &  $\Phi_{123} - \Phi_{213} + \Phi_{314}$\\
  $G_{1g}$  & 2 &  $\Phi_{124} - \Phi_{214} + \Phi_{324}$  \\ \hline
  $G_{1g}$  & 1 &  $2\Phi_{114} + 2\Phi_{332} - \Phi_{123} -
    \Phi_{213} + 2\Phi_{134} - \Phi_{314}$ \\ 
  $G_{1g}$  & 2 &  $-2\Phi_{223} - 2\Phi_{441} + \Phi_{124} -
    \Phi_{214} - 2\Phi_{234} + \Phi_{324}$  \\ \hline
  $G_{1u}$  & 1 &  $\Phi_{112} - \Phi_{334}$  \\ 
  $G_{1u}$  & 2 &  $-\Phi_{221} + \Phi_{443}$  \\ \hline
  $G_{1u}$  & 1 &  $\Phi_{123} - \Phi_{213} - \Phi_{314}$  \\
  $G_{1u}$  & 2 &  $\Phi_{124} - \Phi_{214} - \Phi_{324}$  \\ \hline
  $G_{1u}$  & 1 &  $2\Phi_{114} - 2\Phi_{332} - \Phi_{123} -
    \Phi_{213} - 2\Phi_{134} + \Phi_{314}$  \\ 
  $G_{1u}$  & 2 &  $-2\Phi_{223} + 2\Phi_{441} + \Phi_{124} +
    \Phi_{214} + 2\Phi_{234} - \Phi_{324}$  \\ \hline
  $H_{g}$   & 1 &  $\sqrt{3}(\Phi_{113} + \Phi_{331})$\\
  $H_{g}$   & 2 &  $\Phi_{114} + \Phi_{332} + \Phi_{123} +
    \Phi_{213} - 2\Phi_{134} + \Phi_{314}$  \\
  $H_{g}$   & 3 &  $\Phi_{223} - \Phi_{441} + \Phi_{124} +
    \Phi_{214} - 2\Phi_{234} + \Phi_{324}$  \\
  $H_{g}$   & 4 &  $\sqrt{3}(\Phi_{224} + \Phi_{442})$ \\ \hline
  $H_{u}$   & 1 &  $\sqrt{3}(\Phi_{113} - \Phi_{331})$\\ 
  $H_{u}$   & 2 &  $\Phi_{114} - \Phi_{332} + \Phi_{123} +
    \Phi_{213} + 2\Phi_{134} - \Phi_{314}$ \\
  $H_{u}$   & 3 &  $\Phi_{223} - \Phi_{441} + \Phi_{124} +
    \Phi_{214} + 2\Phi_{234} - \Phi_{324}$  \\
  $H_{u}$   & 4 &  $\sqrt{3}(\Phi_{224} - \Phi_{442})$ \\ \hline
\end{tabular}
\end{center}
\end{table}


\section{Survey of Results}

Now that we have provided a detailed description of the procedure for
the calculation of excited baryons on the lattice, we will now present
an overview of recent lattice calculations of the excited baryon
spectrum.  The emphasis of most calculations has been the
demonstration that the excited nucleon spectrum is indeed accessible
to lattice calculations, and most of the calculations share several
features.
Namely, they are in the quenched approximation to QCD, are obtained
on lattice volumes of $2 - 3~{\rm fm}$, employ pseudoscalar masses of
around $500~{\rm MeV}$, and make fairly limited investigations of the
systematic uncertainties on the calculations.  

The most comprehensive study of the excited nucleon spectrum has been
obtained using FLIC fermions, detailed in Refs.~\cite{CSSM,CSSM32},
and we will describe these calculations at length in these lectures.
For some of the lowest-lying resonances, such as the Roper resonance
and the odd-parity partner of the nucleon, there have been studies of the
systematic uncertainties in the calculations, in particular arising
from the finite lattice spacing, and the need to perform a chiral
extrapolation \cite{Morel:2003fj}.

For completeness, we will first briefly describe the gauge and fermion
actions used in the FLIC fermion analysis.
Additional details of the simulations can be found in
Refs.~\cite{FATJAMES,CSSM,CSSM32}.

\subsection{Lattice Actions for FLIC Calculation}

For the gauge fields, a mean-field improved plaquette plus rectangle
action is used. 
The simulations are performed on a $16^3\times 32$ lattice at $\beta=4.60$,
which corresponds to a lattice spacing of $a = 0.122(2)$~fm set by a
string tension analysis incorporating the lattice Coulomb potential
\cite{Edwards:1998xf} with $\sqrt\sigma = 440$~MeV.
For the quark fields, a Fat-Link Irrelevant Clover
(FLIC) \cite{FATJAMES} action is implemented. A detailed description
of the FLIC fermion action can be found in the contribution by Zanotti
{\it et al.} in this series.

A fixed boundary condition in the time direction is used for the fermions
by setting $U_t(\vec x, N_t) = 0\ \forall\ \vec x$ in the hopping terms
of the fermion action, with periodic boundary conditions imposed in the
spatial directions.
Gauge-invariant Gaussian smearing \cite{Gusken:qx} in the spatial
dimensions is applied at the source to increase the overlap of the
interpolating operators with the ground states.

Five masses are used in the calculations \cite{FATJAMES} and the strange 
quark mass is taken to be the second heaviest quark mass in each case.
The analysis is based on a sample of 400 configurations, and the error
analysis is performed by a third-order, single-elimination jackknife,
with the $\chi^2$ per degree of freedom ($N_{\rm DF}$) obtained via
covariance matrix fits.

\subsection{Spin-$\frac{1}{2}$ Baryons}

\begin{figure}[t!]
\begin{center}
\includegraphics[height=\hsize,angle=270]{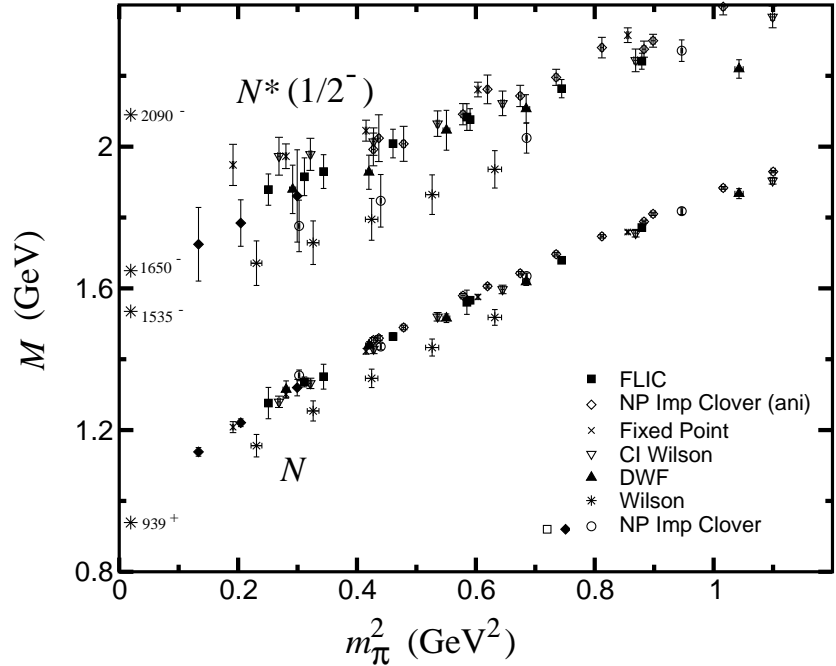} 
\caption{Masses of the nucleon ($N$) and the lowest $J^P={1\over 2}^-$
  excitation (``$N^*$'').  The FLIC and Wilson results are from
  Ref~\protect\cite{CSSM}, the DWF \protect\cite{DWF}, Fixed Point
  \protect\cite{FP}, Chirally Improved (CI) Wilson \protect\cite{FP},
  NP improved clover \protect\cite{RICHARDS} and NP improved anisotropic
  clover \protect\cite{Nakajima:2001js} results are shown for
  comparison.  The empirical nucleon and low lying $N^*({1\over 2}^-)$
  masses are indicated by the asterisks along the ordinate.
\label{nstar}}
\end{center}
\end{figure}

%

In Fig.~\ref{nstar} we show the nucleon and $N^*({1\over 2}^-)$ masses
as a function of the pseudoscalar meson mass squared, $m_\pi^2$.  The
filled squares indicate the results for
the FLIC action, and the stars for the Wilson action \cite{CSSM} (the
Wilson points are obtained from a sample of 50 configurations).
We note here that the spatial size of the lattice
is $L=1.95$~fm and that from the values of $m_\pi$ given in
Table~1 of Ref.~\cite{CSSM} we have $m_\pi L \geq 5.52$.

For comparison, we also show results from earlier simulations with
domain wall fermions (DWF) \cite{DWF} (filled triangles), a
nonperturbatively (NP) improved clover action on anisotropic lattices
at three different lattice spacings \cite{Nakajima:2001js} (open
diamonds), an NP improved clover action at $\beta=6.2$ \cite{RICHARDS}
(open squares, open circles and filled diamonds),
and results using Fixed Point (FP) (crosses)
and Chirally Improved (CI) (open inverted triangles) \cite{FP} fermion
actions.
The scatter of the different NP improved results is due to different
source smearing and volume effects: the open squares are obtained by
using fuzzed sources and local sinks, the open circles use Jacobi
smearing at both the source and sink, while the filled diamonds, which
extend to smaller quark masses, are obtained from a larger lattice
($32^3 \times 64$) using Jacobi smearing.
The empirical masses of the nucleon and the three lowest ${1\over 2}^-$
excitations are indicated by the asterisks along the ordinate.
In an unquenched calculation, the results may shift by the order of
$10\%$ compared with a quenched calculation \cite{YOUNG}.

There is excellent agreement between the different improved actions
for the nucleon mass, in particular between the FLIC \cite{CSSM}, DWF
\cite{DWF}, NP improved clover \cite{Nakajima:2001js,RICHARDS}, FP and
CI \cite{FP} results.  On the other hand, the Wilson results lie
systematically low in comparison to these due to the large ${\cal O}
(a)$ errors in this action \cite{FATJAMES}.
A similar pattern is repeated for the $N^*({1\over 2}^-)$ masses.
Namely, the FLIC, DWF, NP improved clover, FP and CI masses are in good
agreement with each other, while the Wilson results again lie
systematically lower.
A mass splitting of around 400~MeV is clearly visible between the $N$
and $N^*$ for all actions, including the Wilson action, despite its
poor chiral properties.  Furthermore, the trend of the $N^*({1\over
2}^-)$ data with decreasing $m_\pi$ is consistent with the approach to
the mass of the lowest-lying physical negative parity $N^*$ states.

\begin{figure}[t]
\begin{center}
\leavevmode
\epsfig{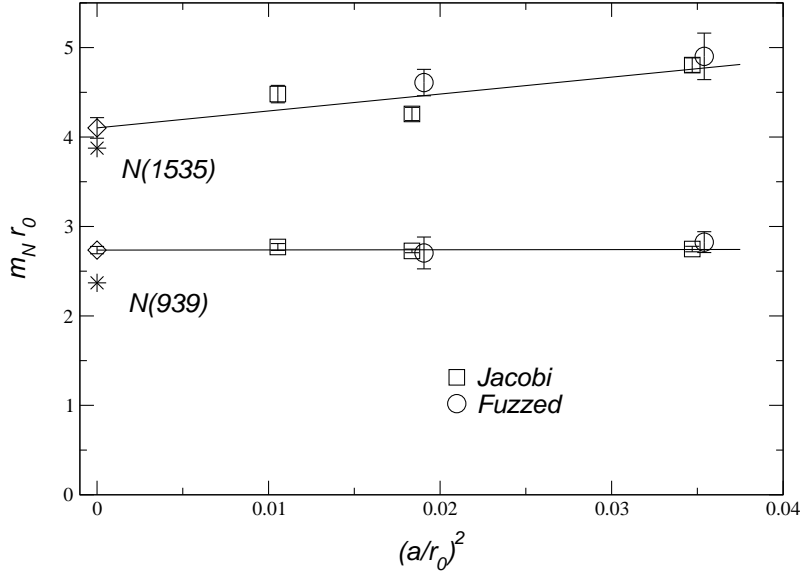}
\vspace*{0.5cm}
\caption{ The masses of the lowest-lying positive- and negative-parity
baryons in units of $r_0^{-1}$ versus $a^2$ in units of $r_0^2$, where
$r_0$ is Sommer's scale \protect\cite{SOMMER}.  The
lines are linear fits in $a^2/r_0^2$ to the positive- and
negative-parity baryon masses; different plotting symbols correspond
to different types of smearing.  Also shown are the physical
values.}\label{fig:continuum_extrap}
\end{center}
\end{figure}

In the case of the NP-improved clover fermion action,
with ${\cal O}(a^2)$ discretisation errors, the calculation has been
performed at three values of the lattice spacing, enabling a continuum
extrapolation to be performed \cite{RICHARDS}.  This is shown in
Fig.~\ref{fig:continuum_extrap}, although it should be noted that a
simple linear chiral extrapolation was performed in this calculation.
There is a suggestion of a somewhat larger lattice spacing dependence
for the higher excited resonance, emphasising the need to perform a
careful analysis of systematic uncertainties in future calculations.

\begin{figure}[t!] 
\begin{center}
\leavevmode
\epsfig{figure=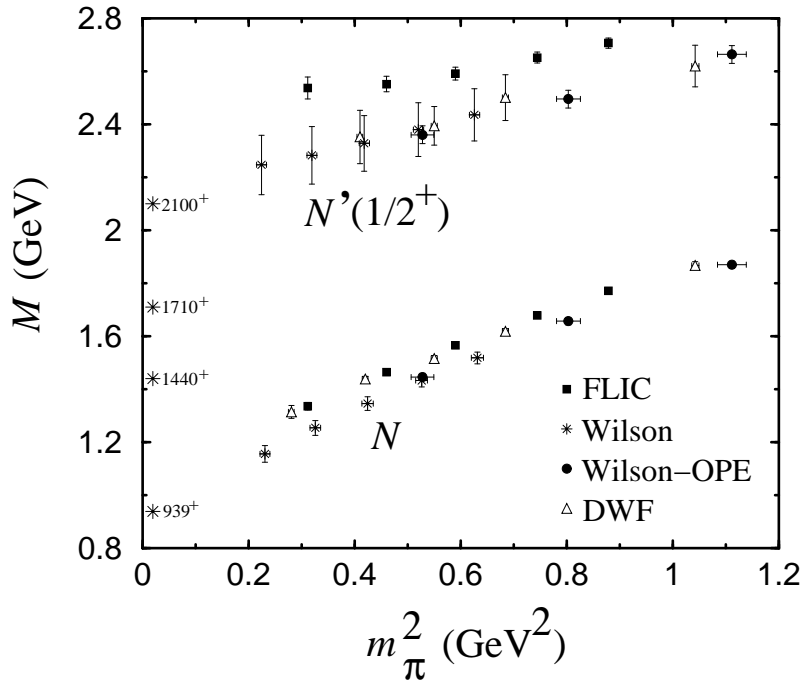,height=9cm}
\vspace*{0.5cm}
\caption{Masses of the nucleon, and the lowest $J^P={1\over 2}^+$
  excitation (``$N'$'').  The FLIC results \protect\cite{CSSM} are
  compared with the DWF \protect\cite{DWF} and Wilson-OPE
  \protect\cite{DEREK} analyses, as well as with the Wilson results
  from Ref.~\protect\cite{CSSM}.  The empirical nucleon and low lying
  $N^*({1\over 2}^+)$ masses are indicated by asterisks, with physical
  masses given in MeV.
\label{nprime}}
\end{center}
\end{figure}

Figure~\ref{nprime} shows the mass of the $J^P = {1\over 2}^+$ states
(the excited state is denoted by ``$N'(1/2^+)$'').
As is long known, the positive parity $\chi_2$ interpolating field
does not have good overlap with the nucleon ground state \cite{DEREK}
and a correlation matrix analysis confirms this result \cite{CSSM},
as discussed below.  It has been speculated that $\chi_2$ may have
overlap with the lowest ${1\over 2}^+$ excited state, the $N^*(1440)$
Roper resonance \cite{DWF}.  In addition to the FLIC and Wilson
results from the present analysis, we also show in Fig.~\ref{nprime}
the DWF results \cite{DWF}, and those from an earlier analysis with
Wilson fermions together with the operator product expansion (OPE)
\cite{DEREK}.  The physical values of the lowest three ${1\over 2}^+$
excitations of the nucleon are indicated by the asterisks.

The most striking feature of the data is the relatively large excitation
energy of the $N'({1\over 2}^+)$, some 1~GeV above the nucleon.
There is little evidence, therefore, that this state is the $N^*(1440)$
Roper resonance.
While it is possible that the Roper resonance may have a strong nonlinear
dependence on the quark mass at $m_\pi^2 \sim 0.2$~GeV$^2$, arising from,
for example, pion loop corrections, it is unlikely that this behaviour
would be so dramatically different from that of the $N^*(1535)$ so as
to reverse the level ordering obtained from the lattice.
A more likely explanation is that the $\chi_2$ interpolating field does
not have good overlap with either the nucleon or the $N^*(1440)$, but
rather (a combination of) excited ${1\over 2}^+$ state(s).

Recall that in a constituent quark model in a harmonic oscillator basis,
the mass of the lowest mass state with the Roper quantum numbers is
higher than the lowest $P$-wave excitation. It seems that neither the
lattice data (at large quark masses and with our interpolating fields)
nor the constituent quark model have good overlap with the Roper
resonance.
Better overlap with the Roper is likely to require more exotic
interpolating fields.

As mentioned in Section~\ref{history}, Lee {\em et al.}
\cite{BAYESIAN} have performed a calculation using overlap fermions
with pion masses down to $\sim 180$~MeV. Using new constrained
curve fitting techniques, they extract excited states from
a single correlation function calculated with the standard nucleon
interpolating field in Eq.~(\ref{chi1p}). The results from this
calculation exhibit a dramatic drop in the mass of the first excited
${1\over 2}^+$ state of the nucleon at light pion masses, reversing
the level ordering of the first ${1\over 2}^+$ and ${1\over 2}^-$
excited states.
It is important, however, that this result be shown to
be independent of the constrained
curve fitting techniques adopted for this analysis.
A correlation matrix analysis involving
several operators would shed considerable light on this issue.

Recently there has also been speculation that the Roper resonance suffers
from large finite volume errors \cite{Sasaki:2003xc}. To study this
issue, the authors of Ref.~\cite{Sasaki:2003xc} calculate the nucleon
and its first positive and negative parity excitations on three
different lattice volumes ($La=1.5,\,2.2$ and 3.0 fm).
Using Maximum Entropy Methods, they find that on large volume lattices
($\simge 3.0$ fm) the mass of the ${1\over 2}^+$ excited nucleon state
is reduced.  A similar analysis remains to be performed for the first
${1\over 2}^-$ nucleon state obtained from the same fermion action.
At present Wilson fermion scaling violations allow the ${1\over 2}^+$
excited nucleon state to sit lower than the first ${1\over 2}^-$
nucleon state obtained from the improved DWF action.
It is essential to compare the masses of these states using the same
fermion actions to remove systematic errors such as these.  Similarly,
a correlation matrix analysis involving several operators remains
desirable.

The BGR Collaboration has performed a calculation of the excited
nucleon spectrum on two lattice volumes ($La \simeq 1.8$ and 2.4 fm)
using Fixed Point and Chirally Improved Wilson fermions \cite{FP}.
Using three different operators to create the nucleon states, the
lowest two states in both the ${1\over 2}^+$ and ${1\over 2}^-$
channels are extracted using a $3\times3$ correlation matrix. The
eigenvectors and eigenvalues determined from the correlation matrix
analysis should correspond to the solutions of \eqn{eq:corr_ev}.  It
is unclear, however, whether the constraints which have been employed
in Ref.~\cite{FP} are physical, and in particular whether the
eigenvectors correspond to the optimal projections onto physical
states defined by the eigenvectors $u$ of \eqn{eq:corr_ev}. With this
caveat, a splitting between the two ${1\over 2}^-$ states is
identified and the excited ${1\over 2}^+$ state is observed to sit
above the ${1\over 2}^-$ states, in agreement with Ref.~\cite{CSSM} (see next
section). The results on two volumes do not exhibit any volume
dependence.

%
%

In the next section we present results from a $2\times2$ correlation
matrix analysis using FLIC fermions on a single lattice volume.




\subsection{Resolving the resonances}
\label{resolve}

\begin{figure}[t] 
\begin{center}
\epsfig{figure=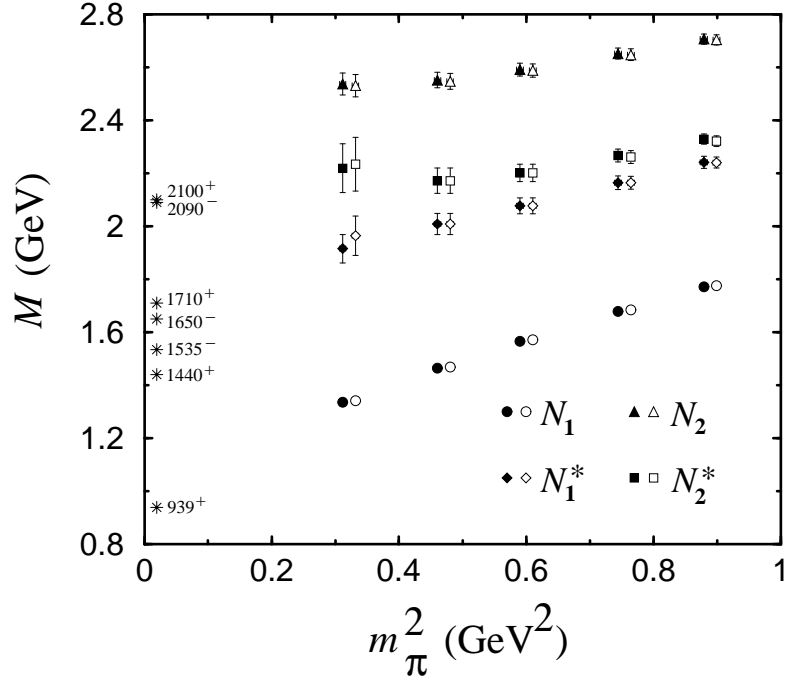,height=9cm}
\vspace*{0.5cm}
\caption{Masses of the $J^P = {1\over 2}^+$ and ${1\over 2}^-$ nucleon
  states, for the FLIC action \protect\cite{CSSM}.  The positive
  (negative) parity states
  are labeled $N_1$ ($N_1^*$) and $N_2$ ($N_2^*$).  The results from
  the projection of the correlation matrix are shown by the filled
  symbols, whereas the results from the standard fits to the
  $\chi_1\overline\chi_1$ and $\chi_2\overline\chi_2$ correlation
  functions are shown by the open symbols (offset to the right for
  clarity).  Empirical masses of the low lying ${1 \over 2}^\pm$
  states are indicated by the asterisks.
\label{nchipi2}}
\end{center}
\end{figure}

The mass splitting between the two lightest $N^*({1\over 2}^-)$
states ($N^*(1535)$ \& $N^*(1650)$) can be studied by considering the
odd parity content of the $\chi_1$ and $\chi_2$ interpolating fields in
Eqs.~(\ref{chi1p}) and (\ref{chi2p}).
Recall that the ``diquarks'' in $\chi_1$ and $\chi_2$ couple differently
to spin, so that even though the correlation functions built up from the
$\chi_1$ and $\chi_2$ fields will be made up of a mixture of many excited
states, they will have dominant overlap with different states \cite{DEREK,LL}.
By using the correlation-matrix techniques described in
Ref.~\cite{CSSM} (see also Appendix),
two separate mass states are extracted from the $\chi_1$ and $\chi_2$
interpolating fields.
The results from the correlation matrix analysis are shown by the
filled symbols in Fig.~\ref{nchipi2}, and are compared to the standard
``naive'' fits performed directly on the diagonal correlation functions,
$\chi_1 \overline{\chi}_1$ and $\chi_2 \overline{\chi}_2$, indicated by
the open symbols.

The results indicate that indeed the $N^*({1\over 2}^-)$ largely
corresponding to the $\chi_2$ field (labeled ``$N_2^*$'') lies above
the $N^*({1\over 2}^-)$ which can also be isolated via Euclidean time
evolution with the $\chi_1$ field (``$N_1^*$'') alone.
The masses of the corresponding positive parity states, associated with
the $\chi_1$ and $\chi_2$ fields (labeled ``$N_1$'' and ``$N_2$'',
respectively) are shown for comparison.
For reference, we also list the experimentally measured values of the
low-lying ${1\over 2}^\pm$ states.
It is interesting to note that the mass splitting between the positive
parity $N_1$ and negative parity $N_{1,2}^*$ states (roughly 400--500~MeV)
is similar to that between the $N_{1,2}^*$ and the positive parity $N_2$
state, reminiscent of a constituent quark--harmonic oscillator picture.

\begin{figure}[p] 
\begin{center}
\epsfig{figure=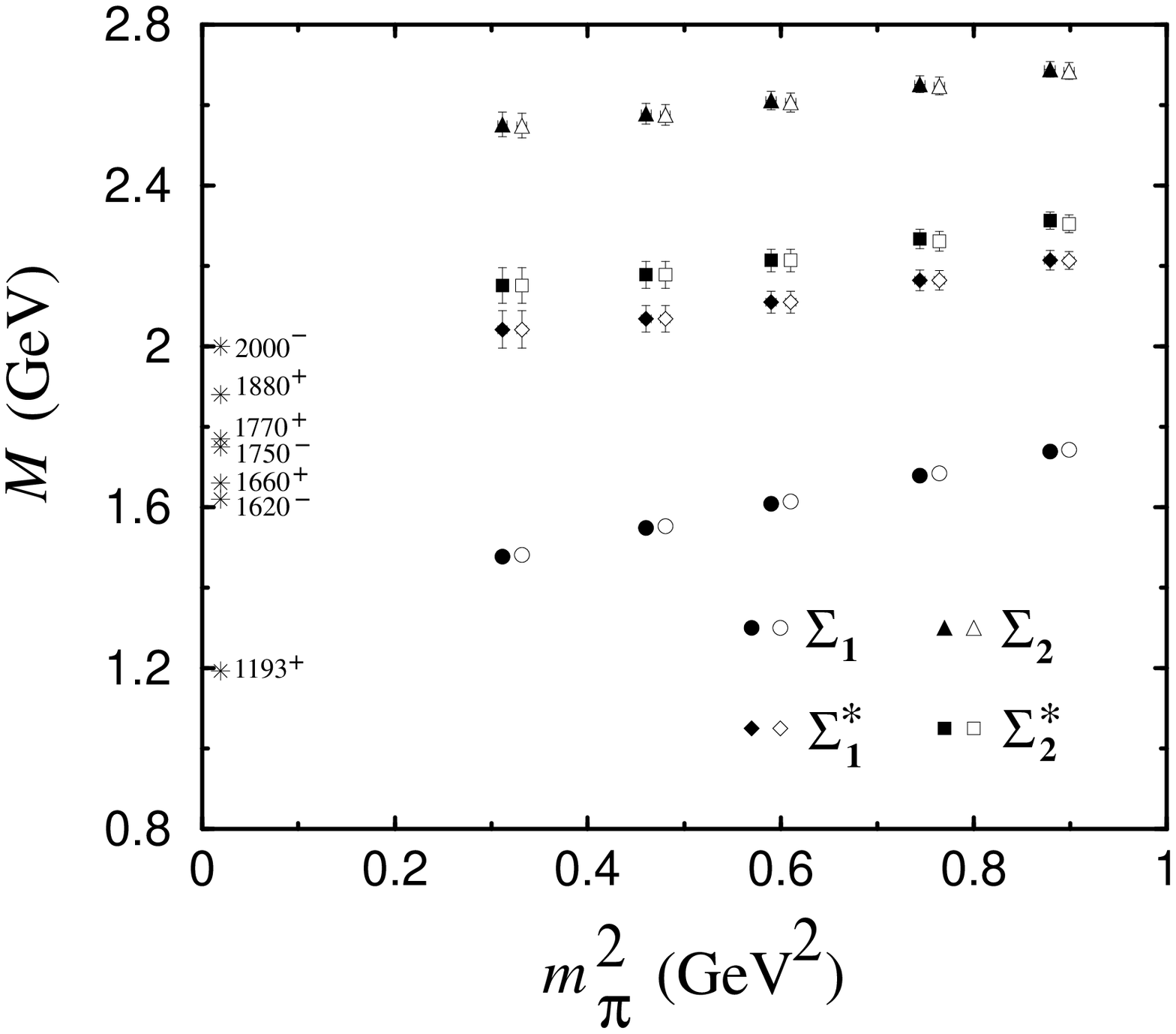,height=7.5cm}
\vspace*{0.23cm}
\caption{As in Fig.~\ref{nchipi2} but for the $\Sigma$ baryons.
\label{schi}}
\vspace*{0.45cm}
\epsfig{figure=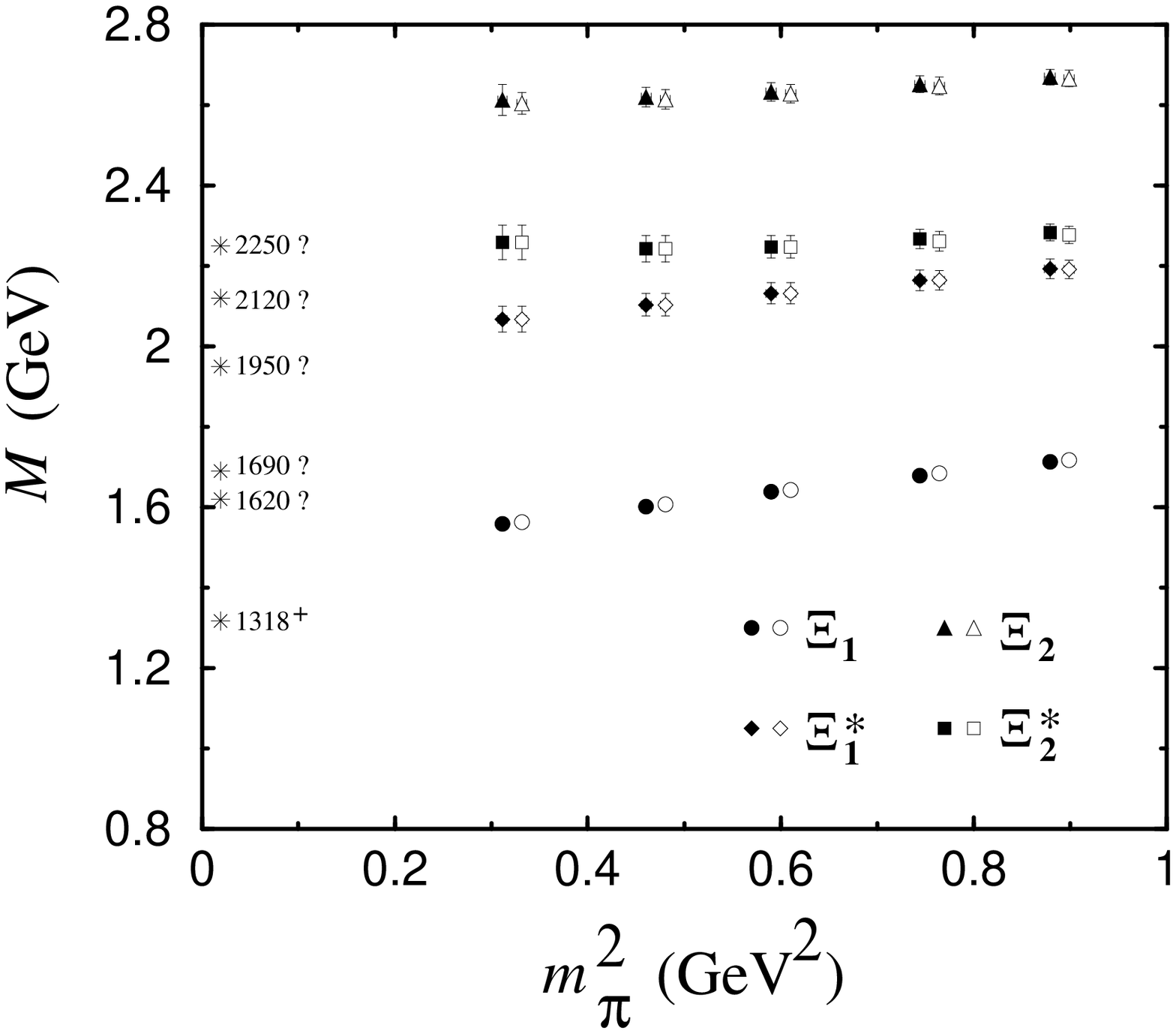,height=7.5cm}
\vspace*{0.23cm}
\caption{As in Fig.~\ref{nchipi2} but for the $\Xi$ baryons.
        The $J^P$ values of the excited states marked with
        ``?'' are undetermined.
        \label{Xifig}}
\end{center}
\end{figure}

Turning to the strange sector, in Fig.~\ref{schi} we show the masses of the
positive and negative parity $\Sigma$ baryons calculated from the FLIC
action \cite{CSSM} compared with the physical masses of the known positive and
negative parity states.
The pattern of mass splittings is similar to that found in
Fig.~\ref{nchipi2} for the nucleon.
Namely, the ${1\over 2}^+$ state associated with the $\chi_1$ field
appears consistent with the empirical $\Sigma(1193)$ ground state,
while the ${1\over 2}^+$ state associated with the $\chi_2$ field lies
significantly above the observed first (Roper-like) ${1\over 2}^+$
excitation, $\Sigma^*(1660)$.
There is also evidence for a mass splitting between the two negative parity
states, similar to that in the nonstrange sector.
%

The spectrum of the strangeness --2 positive and negative parity $\Xi$
hyperons is displayed in Fig.~\ref{Xifig}.  Once again, the pattern of
calculated masses repeats that found for the $\Sigma$ and $N$ masses
in Figs.~\ref{nchipi2} and \ref{schi}, and for the respective coupling
coefficients \cite{CSSM}.
The empirical masses of the physical $\Xi^*$ baryons are denoted by
asterisks. 
However, for all but the ground state $\Xi(1318)$, the $J^P$
values are not known.

\begin{figure}[p] 
\begin{center}
\epsfig{figure=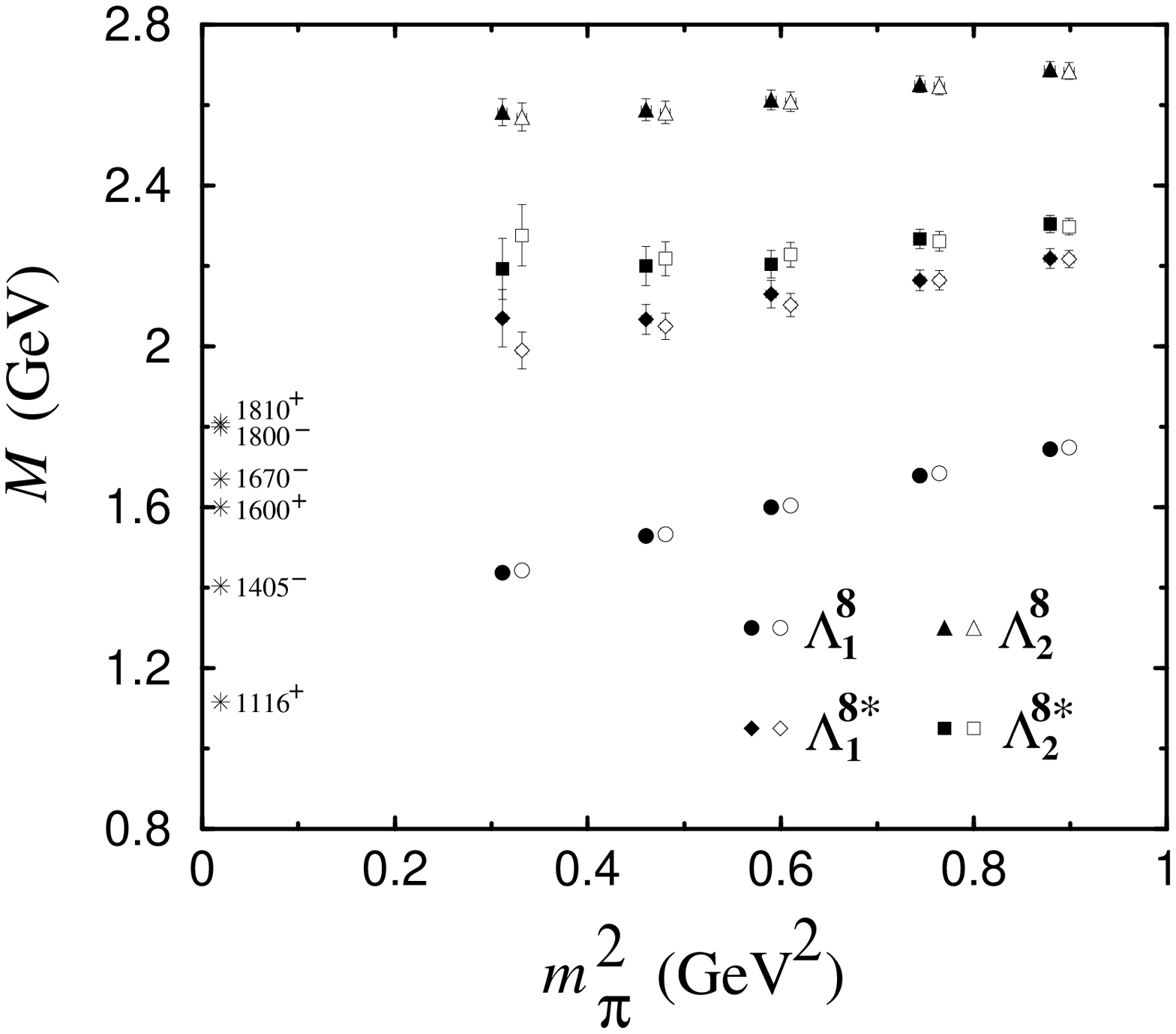,height=7.5cm}
\vspace*{0.23cm}
\caption{As in Fig.~\ref{nchipi2} but for the $\Lambda$ states obtained using the
        $\Lambda^8$ interpolating field.
        \label{l8chipi2}}
\vspace*{0.45cm}
\epsfig{figure=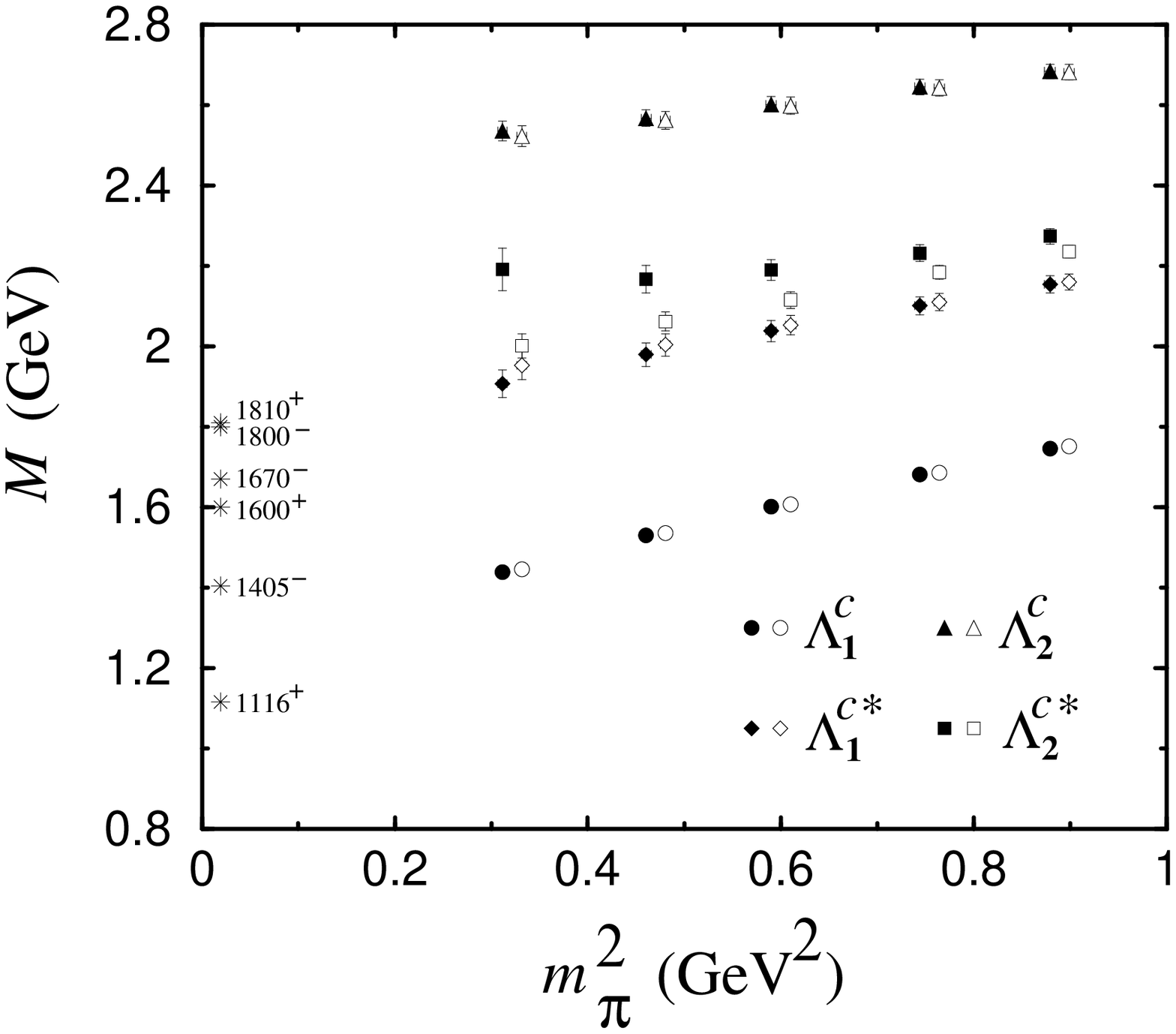,height=7.5cm}
\vspace*{0.23cm}
\caption{As in Fig.~\ref{nchipi2} but for the $\Lambda$ states obtained using the
        $\Lambda^c$ interpolating field.
        \label{lcchipi2}}
\end{center}
\end{figure}

\begin{figure}[t] 
\begin{center}
\epsfig{figure=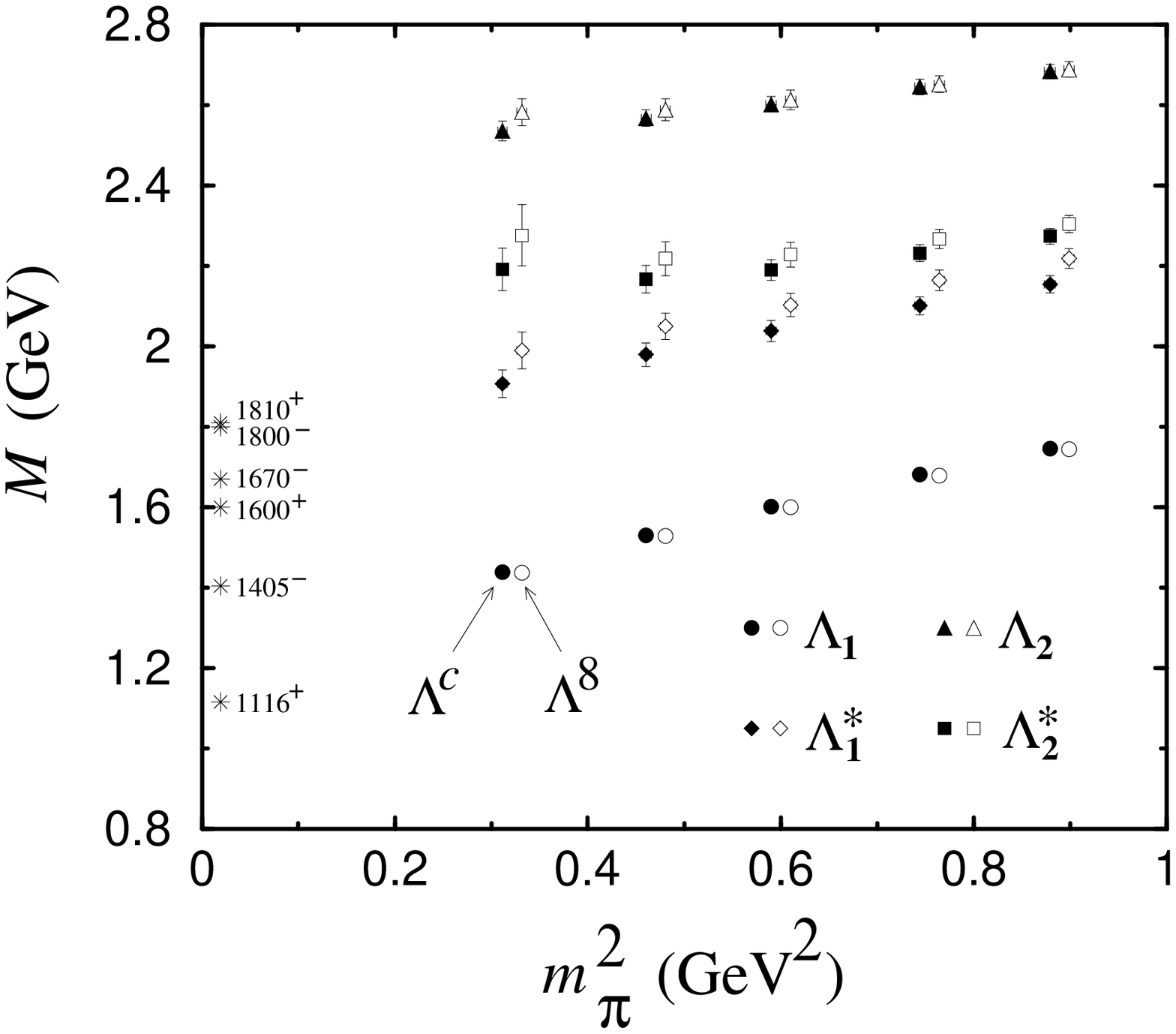,height=9cm}
\vspace*{0.5cm}
\caption{Masses of the positive and negative parity $\Lambda$ states, for
        the octet $\Lambda^8$ (open symbols) and ``common'' $\Lambda^c$
        (filled symbols) interpolating fields with the FLIC action
	\protect\cite{CSSM}.
        The positive (negative) parity states labeled $\Lambda_1$
        ($\Lambda_1^*$) and $\Lambda_2$ ($\Lambda_2^*$) are the two
        states obtained from the correlation matrix analysis of
        the $\chi_1^\Lambda$ and $\chi_2^\Lambda$ interpolating
        fields.  Empirical masses of the low lying
        ${1\over 2}^\pm$ states are indicated by the asterisks.
\label{lchiCM}}
\end{center}
\end{figure}

Finally, we consider the $\Lambda$ hyperons.  In Figs.~\ref{l8chipi2}
and \ref{lcchipi2} we compare results obtained from the $\Lambda^8$
and $\Lambda^c$ interpolating fields, respectively, using the two
different techniques for extracting masses. 
A direct comparison between the positive and negative parity masses for
the $\Lambda^8$ (open symbols) and $\Lambda^c$ (filled symbols) states
extracted from the correlation matrix analysis, is shown in
Fig.~\ref{lchiCM}.
A similar pattern of mass splittings to that for the
$N^*$ spectrum of Fig.~8 is observed.
In particular, the negative parity $\Lambda_1^*$ state (diamonds) lies
$\sim 400$~MeV above the positive parity $\Lambda_1$ ground state
(circles), for both the $\Lambda^8$ and $\Lambda^c$ fields.
There is also clear evidence of a mass splitting between the 
$\Lambda_1^*$ (diamonds) and $\Lambda_2^*$ (squares).

Using the naive fitting scheme (open symbols in Figs.~\ref{l8chipi2}
and \ref{lcchipi2}) misses the mass splitting between $\Lambda_1^*$ and
$\Lambda_2^*$ for the ``common'' interpolating field.
Only after performing the correlation matrix analysis is it possible
to resolve two separate mass states, as seen by the filled symbols in
Fig.~\ref{lcchipi2}.
As for the other baryons, there is little evidence that the $\Lambda_2$
(triangles) has any significant overlap with the first positive parity
excited state, $\Lambda^*(1600)$ (cf. the Roper resonance, $N^*(1440)$,
in Fig.~\ref{nchipi2}).

While it seems plausible that nonanalyticities in a chiral extrapolation
\cite{MASSEXTR} of $N_1$ and $N_1^*$ results could eventually lead to
agreement with experiment, the situation for the $\Lambda^* (1405)$ is
not as compelling. 
Whereas a 150~MeV pion-induced self energy is required for the 
$N_1 ,\ N_1^*$ and $\Lambda_1$, 400~MeV is required to approach the 
empirical mass of the $\Lambda^* (1405)$.
This may not be surprising for the octet fields, as the $\Lambda^*(1405)$, 
being an SU(3) flavour singlet, may not couple strongly to an SU(3) octet 
interpolating field. 
Indeed, there is some evidence of this in Fig.~\ref{lchiCM}. 
This large discrepancy of 400~MeV suggests that relevant physics giving rise to a
light $\Lambda^* (1405)$ may be absent from 
simulations in the quenched approximation.
The behaviour of the $\Lambda_{1,2}^*$ states may be modified at small
values of the quark mass through nonlinear effects associated with
Goldstone boson loops including the strong coupling of the
$\Lambda^*(1405)$ to $\Sigma\pi$ and $\bar K N$ channels.
While some of this coupling will survive in the quenched approximation,
generally the couplings are modified and suppressed \cite{YOUNG,SHARPE}.
It is also interesting to note that the $\Lambda_1^*$ and
$\Lambda_2^*$ masses display a similar behaviour to that seen for the
$\Xi_1^*$ and $\Xi_2^*$ states, which are dominated by the heavier
strange quark.
Alternatively, the study of more exotic interpolating fields may indicate
the the $\Lambda^*(1405)$ does not couple strongly to $\chi_1$ or
$\chi_2$.
Investigations at lighter quark masses involving quenched chiral
perturbation theory will assist in resolving these issues.

\subsection{Spin-$\frac{3}{2}$ Baryons}
\label{32results}

\begin{figure}[t]
\begin{center}
\includegraphics[height=0.85\hsize,angle=90]{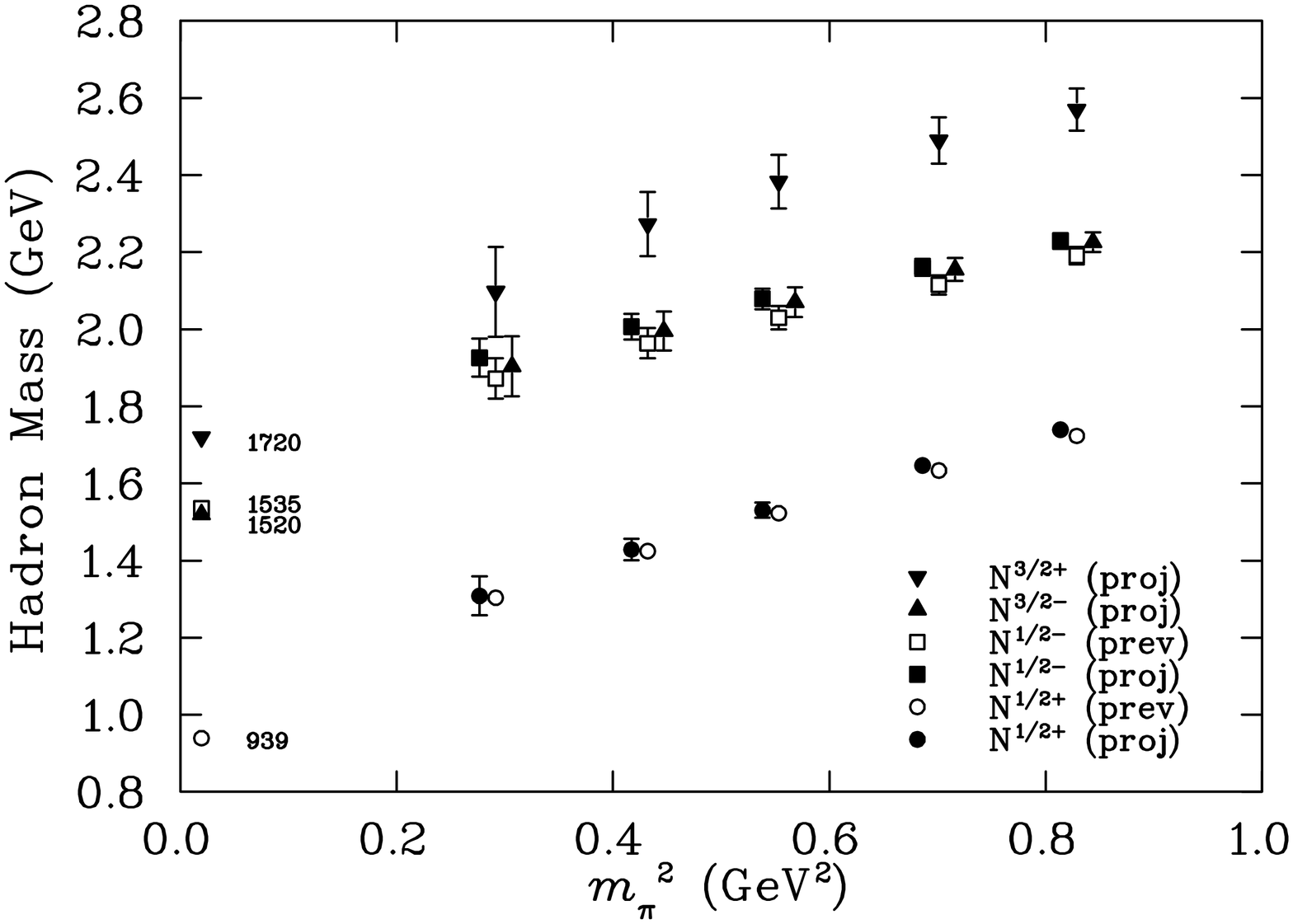} 
\vspace*{0.5cm}
\caption{Masses of the spin projected $N{3\over 2}^-$ (filled triangles),
  $N{3\over 2}^+$ (filled inverted triangles), $N{1\over 2}^+$
  (filled circles), and $N{1\over 2}^-$ (filled squares)
  isospin-${1\over 2}$ states \protect\cite{CSSM32}. For comparison,
  previous results from the direct calculation of the
  $N{1\over 2}^+$ (open circles) and $N{1\over 2}^-$ (open squares)
  from Fig.~\protect\ref{nstar} are also shown. The empirical values
  of the masses of the $N{1\over 2}^+\,(939)$, $N{1\over 2}^-\,
  (1535)$, $N{3\over 2}^-\, (1520)$ and $N{3\over 2}^+\, (1720)$  are
  indicated (in MeV) on the left-hand-side at the physical pion mass.
\label{Nvsmpi2}}
\end{center}
\end{figure}

The interpolating field defined in Eq.~(\ref{N32IF}) has overlap with
spin-$\frac{1}{2}$ and spin-$\frac{3}{2}$ states of both parities.
After performing appropriate spin projections on the correlation
functions, the masses of the $N{3\over 2}^\pm$ and $N{1\over 2}^\pm$
states are extracted and displayed in Fig.~\ref{Nvsmpi2} as a
function of $m_\pi^2$.
Earlier results for the $N\frac{1}{2}^\pm$ states using the standard
spin-$1\over 2$ interpolating field \cite{FATJAMES,CSSM} from
Eq.~(\ref{chi1p}) are also shown with open symbols in
Fig.~\ref{Nvsmpi2} for reference.
It is encouraging to note the agreement between the
spin-projected ${1\over 2}^\pm$ states obtained from the spin-${3\over 2}$
interpolating field in Eq.~(\ref{N32IF}) and the earlier ${1\over 2}^\pm$
results from the same gauge field configurations.
We also observe that the $N\frac{3}{2}^-$ state has approximately the
same mass as the spin-projected $N\frac{1}{2}^-$ state which is
consistent with the experimentally observed masses.
The results for the $N\frac{3}{2}^-$ state in Fig.~\ref{Nvsmpi2}
indicate a clear mass splitting between the $N{3\over 2}^+$ and
$N{3\over 2}^-$ states obtained from the spin-${3\over 2}$
interpolating field, with a mass difference around 300~MeV. This is
slightly larger than the experimentally observed mass difference of
200~MeV.



\begin{figure}[t]
\begin{center}
\includegraphics[height=0.85\hsize,angle=90]{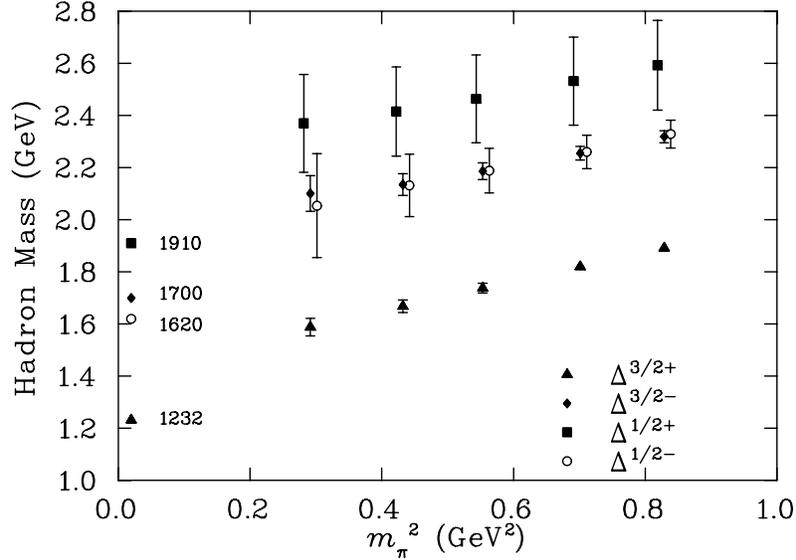} 
\vspace*{0.5cm}
\caption{Masses of the spin-projected $\Delta{3\over 2}^\pm$ and
  $\Delta{1\over 2}^\pm$ isospin-${3\over 2}$ resonances
  \protect\cite{CSSM32}.
  The empirical values of the
  masses of the $\Delta^{3/2^+}(1232)$, $\Delta{3\over 2}^-(1700)$,
  $\Delta{1\over 2}^-(1620)$ and $\Delta{1\over 2}^+(1910)$ are
  shown (in MeV) on the left-hand-side at the physical pion mass.
  \label{Dpi2}}
\end{center}
\end{figure}

Turning now to the isospin-${3\over 2}$ sector, 
results for the $\Delta{3\over 2}^+$ and $\Delta{3\over 2}^-$
masses are shown in Fig.~\ref{Dpi2} as a function of $m_\pi^2$.
The trend of the $\Delta{3\over 2}^+$ data points with decreasing $m_{\pm}$
is clearly towards the $\Delta(1232)$, although some nonlinearity with
$m_\pi^2$ is expected near the chiral limit \cite{MASSEXTR,YOUNG}.
The mass of the $\Delta{3\over 2}^-$ lies some 500~MeV above that of
its parity partner, although with somewhat larger errors. 

After performing a spin projection to extract the $\Delta{1\over
  2}^\pm$ states a discernible, but noisy, signal is detected. This
indicates that the interpolating field in Eq.~(\ref{deltaIF}) has only
a small overlap with spin-$\frac{1}{2}$ states.
However, with 400 configurations we are able to extract a mass for the
spin-$\frac{1}{2}$ states at early times, shown in
Fig.~\ref{Dpi2}. Here we see the larger error bars associated with the
$\Delta{1\over 2}^\pm$ states.
The lowest excitation of the ground state, namely the $\Delta{1\over
  2}^-$, has a mass $\sim 350$--400~MeV above the $\Delta{3\over
  2}^+$, with the $\Delta{3\over 2}^-$ possibly appearing heavier.
The $\Delta{1\over 2}^+$ state is found to lie $\sim 100$--200~MeV
above these, although the signal becomes weak at smaller quark masses.
This level ordering is consistent with that observed in the empirical
mass spectrum, which is also shown in the figure.

The $N{1\over 2}^-$ and $\Delta{1\over 2}^-$ states will decay to
$N\pi$ in $S$-wave even 
in the quenched approximation \cite{QQCD}. For all quark masses
considered here, with the possible exception of the lightest quark,
this decay channel is closed for the nucleon.
While there may be some spectral strength in the decay mode,
we are unable to separate it from the resonant spectral strength.

The $N{3\over 2}^+$ and $\Delta{1\over 2}^+$ states will decay to
$N\pi$ in $P$-wave, while $N{3\over 2}^-$ and $\Delta{3\over 2}^-$
states will decay to $N\pi$ in $D$-wave.
Since the decay products of each of these states must then have equal
and opposite momentum and energy given by
$$
E^2 = M^2 + \left( \frac{2\pi}{aL} \right)^2 ,
$$
these states are stable in the present calculations.

\section{Conclusions}

The increasing effort given to the study of the excited baryon and
meson spectrum by the lattice community reflects the appreciation that
the determination of the spectrum provides vital clues to the
dynamics of QCD, and the mechanisms of confinement.  The $N^*$
spectrum in particular has several features, such as the anomalously
light $N^*(1440)$ Roper resonance and the $J^P={1\over 2}^-$
$\Lambda(1405)$, that defy a straightforward interpretation within
the quark model, and whose understanding can help to resolve the
competing pictures of hadron structure.
The impetus to study the $N^*$ resonances has also strengthened
following the observation of the $S = +1$, $\Theta^+$ pentaquark
state, whose properties remain essentially unknown, but whose various
interpretations offer vastly different pictures of baryon structure.

In these lectures, we have described the techniques required to
compute the $N^*$ spectrum in lattice QCD, and given an overview of
the current status of lattice calculations.  The lightest states of
both parities for spin $1/2$ and spin $3/2$ have been successfully
resolved in the quenched approximation to QCD in both the nucleon
(isospin-${1\over 2}$) and $\Delta$ (isospin-${3\over 2}$) sectors,
and in general the level ordering, albeit at relatively large
pseudoscalar masses, $m_{\pi} \ge 500~{\rm MeV}$, is in accord with
that observed experimentally.  At these large pseudoscalar masses
the spectrum largely follows quark-model expectations.

It is important to appreciate that most current spectroscopy
calculations have employed essentially ``$S$-wave'' quark propagators.
%
%
The measurement of a wider basis of interpolating operators will be an
important element of future studies, and the technology to construct
such operators with lattice symmetry properties has now been
developed.  An important by-product of such studies will be insight
into the quark and gluon structure of such hadrons.
 
The realisation that hadronic physics at physical values of the pion
mass is very different from that at $m_{\pi} \ge 300~{\rm MeV}$, and
the consequent need to correctly account for the chiral properties of
the theory, have been some of the most important developments in
lattice QCD of recent years.  FLIC fermions and the development of
fermions having an 
exact analogue of chiral symmetry provide the means to attain such
pion masses.  There are suggestions that such radically different
behaviour at light quark masses has been seen in the excited nucleon
sector.  
With the development of exact dynamical fat-link fermion algorithms
\cite{Kamleh:2004xk,Kamleh:2004ti,Kamleh:2003wb}, FLIC fermions
provide tremendous promise for accessing the light-quark mass regime
of full QCD.

The continuation to physical values of the light quark masses poses
extra challenges to the calculation of the resonance spectrum.  The
excited states are no longer stable under the strong interaction at
sufficiently light quark masses.
Even in the quenched approximation, this instability is manifest
through non-unitary behaviour in the correlators and through
additional non-analytic terms in the chiral expansion of nucleon
masses \cite{Morel:2003fj}.  The
means to study the full spectrum, including the scattering lengths of
multi-particle states, is in principle known, relying on examining the
finite-volume shift in the two-particle spectrum.  However, the method
is computationally very demanding, requiring the measurement of many
operators.
 
With the realisation of large-scale computing facilities expected over
the next several years, we can expect many of these endevours to come
to fruition.
An exciting era for baryon spectroscopy on the lattice lays ahead.

\section*{Acknowledgements}
This work was supported in part by the Australian Research Council and
by DOE contract DE-AC05-84ER40150 under which the Southeastern
Universities Research Association (SURA) operates the Thomas Jefferson
National Accelerator Facility.  Generous grants of supercomputer time
from the Australian Partnership for Advanced Computing (APAC) and the
Australian National Computing Facility for Lattice Gauge Theory are
gratefully acknowledged.

\appendix
\section{Appendix - Correlation Matrix Analysis}

In this section we outline the correlation matrix formalism for
calculations of masses, coupling strengths and optimal interpolating fields.
After demonstrating that the correlation functions are real, we proceed
to show how a matrix of such correlation functions may be used to isolate
states corresponding to different masses, and also to give information 
about the coupling of the operators to each of these states
(see also Ref.~\cite{CSSM}).

\subsection{The $U+U^{\ast}$ method}

A lattice QCD correlation function for the operator $\chi_i
\overline{\chi}_j$, where $\chi_i$ is the \mbox{$i$-th} interpolating
field for a particular baryon (e.g. $\chi_2^{p+}$ in
Section~\ref{spin12IF}), can be written as
\begin{eqnarray}
\label{definition1}
{\cal G}_{ij} &\equiv& \left\langle\Omega|T(\chi_i \overline{\chi}_j)
  |\Omega\right\rangle \\
&=& \frac{\int {\cal D}U {\cal D}\bar\psi {\cal D}\psi
  e^{-S[U,\bar\psi,\psi]} \chi_i
\overline{\chi}_j}
{\int {\cal D}U {\cal D}\bar\psi{\cal D}\psi e^{-S[U,\bar\psi,\psi]}}\,,
\nonumber
\end{eqnarray} 
where spinor indices and spatial coordinates are suppressed for
ease of notation.
The fermion and gauge actions can be separated such that
$ S[U,\bar\psi,\psi]=S_{G}[U] + \bar\psi M[U] \psi $.
Integration over the Grassmann variables $\bar\psi$ and $\psi$ then
gives
\begin{equation}
\label{naive}
{\cal G}_{ij}\ =\ \frac{ \int {\cal D}U e^{-S_{G}[U]} \det(M[U])H_{ij}[U] }
{\int {\cal D}U e^{-S_{G} [U]} \det(M[U])}\ ,
\end{equation}
where the term $H_{ij}$ stands for the sum of all full contractions of
$\chi_i \overline{\chi}_j$.
The pure gauge action $S_G$ and the fermion matrix $M$ satisfy
\begin{eqnarray}
S_G [U] = S_G [U^\ast]\ ,
\label{eqnarray:ActionProperties1}
\end{eqnarray}
and
\begin{eqnarray} 
\widetilde C M[U^{\ast}] \widetilde C^{-1} = M^{\ast}[U]\ ,
\label{eqnarray:ActionProperties2}
\end{eqnarray}
respectively, where $\widetilde C \equiv C\gamma_5$.

Using the result of Eq.~(\ref{eqnarray:ActionProperties2}), one has
\begin{eqnarray}
\det\left(M[U^{\ast}]\right)
 &=& \det\left(M^{\ast}[U]\right)\ , 
\end{eqnarray}
and since ${\rm det}(M[U])$ is real, this leads to
\begin{equation}
\det\left(M[U^{\ast}]\right) = {\det\left(M[U]\right)}\ .
\end{equation}
Thus, $U$ and $U^{\ast}$ are configurations of equal weight in the
measure \\
$\int {\cal D}U {\rm det}(M[U]){\rm exp}\left(-S_{G}[U]\right)$, in which
case ${\cal G}_{ij}$ can be written as
%
\begin{equation}
\label{endrel}
{\cal G}_{ij}
= \frac{1}{2}\left( \frac{\int {\cal D}U e^{-S_{G}[U]} \det(M[U])
\left\{H_{ij}[U] +H_{ij}[U^\ast]\right\} }
{\int {\cal D}U e^{-S_{G}[U]} \det(M[U]) }\right)\, .
\end{equation}
%

Let us define 
\be
G^{\pm}_{ij} \equiv {\rm tr}[ \Gamma_{\pm}
{\cal G}_{ij} ] \, ,
\ee
where ``$\rm tr$'' denotes the spinor trace
and $\Gamma_{\pm}$ is the parity-projection operator defined
in Eq.~(\ref{pProjOp}).
If ${\rm tr} \left[ {\Gamma H_{ij}[U^\ast]} \right]
  = {\rm tr} \left[ {\Gamma H_{ij}^{\ast}[U]} \right]$,
then $G^{\pm}_{ij}$ is real. 
This can be shown by first noting that $H_{ij}$ will be products of
Dirac $\gamma$-matrices, fermion propagators, and link-field operators.
In a $\gamma$-matrix
representation which is Hermitian, such as the Sakurai representation,
$\widetilde C\gamma_\mu \widetilde C^{-1} = \gamma_\mu^*$. 
Fermion propagators have the form $M^{-1}$, and recalling that since
$\widetilde C M[U^{\ast}] \widetilde C^{-1}$=$M^{\ast}[U]$, then we have
$\widetilde C M^{-1}[U^{\ast}] \widetilde C^{-1}$=$(M^{-1}[U])^{\ast}$.
For products of link-field operators $O[U]$ contained in
$H_{ij}$, the condition
$O[U^{\ast}] = O^{\ast}[U]$ is equivalent to the requirement that the
coefficients of all link-products are real.
As long as this requirement is enforced, we can then simply proceed by
inserting $\widetilde C \widetilde C^{-1}$ inside the trace to show that the (spinor-traced)
correlation functions $G^{\pm}_{ij}$ are real.
If one chooses the Dirac
representation, then $\widetilde C\gamma_k \widetilde C^{-1} =
-\gamma_k^*$ and $\widetilde C\gamma_0
\widetilde C^{-1} = \gamma_0^*$. Therefore, in the Dirac
representation of the $\gamma$-matrices, if $H_{ij}$ contains an even number
of spatial $\gamma$-matrices with real coefficients, $G_{ij}^\pm$ is
purely real; otherwise $G_{ij}^\pm$ is purely imaginary.

In summary, the interpolating fields considered here are constructed
using only real coefficients and have no spatial $\gamma$-matrices.
Therefore, the correlation functions $G^{\pm}_{ij}$ are real.
This symmetry is explicitly implemented by 
including both $U$ and $U^*$ in the ensemble averaging
used to construct the lattice
correlation functions, providing an improved unbiased estimator which
is strictly real.
This is easily implemented at the correlation function level by
observing 
$$
M^{-1}(\{U_\mu^*\}) = [C\gamma_5 \, M^{-1}(\{U_\mu\})\,
(C\gamma_5)^{-1}]^*
$$ 
for quark propagators.

\subsection{Recovering Masses, Couplings and Optimal Interpolators}
\label{recover}

Let us again consider the momentum-space two-point function
for $t > 0$,
\begin{equation}
{\cal G}_{ij}(t,\vec{p}) = \sum_{\vec{x}} e^{-i\vec{p} \cdot \vec{x}}
\langle\Omega|
  \chi_i(t,\vec{x}) \overline{\chi}_j(0,\vec 0)
|\Omega\rangle\ .
\end{equation}
At the hadronic level,
\begin{eqnarray*}
{\cal G}_{ij}(t,\vec{p}) &=& \sum_{\vec{x}} e^{-i\vec{p} \cdot \vec{x}}
\sum_{\vec{p}^{ \prime} ,s} \sum_{B}
\langle\Omega| \chi_i(t,\vec{x})| B ,p',s \rangle \nonumber \\
&\times&
\langle B ,p',s | \overline{\chi}_j(0,\vec 0) |\Omega\rangle\ ,
\end{eqnarray*}
where the $|B ,p',s\rangle$ are a complete set of states with momentum
$p'$ and spin $s$
%
\begin{equation}
\sum_{\vec{p}^{\prime}} \sum_{B} \sum_{s}
 |B ,p',s\rangle\langle B ,p',s|=I\ .
\end{equation}
We can make use of translational invariance to write
%
\begin{eqnarray}
{\cal G}_{ij}(t,\vec{p})
&=& \sum_{\vec{x}} e^{-i\vec{p} \cdot \vec{x}}
\sum_{\vec{p}^{\prime}} \sum_{s} \sum_{B}
\left\langle\Omega\left|
  e^{\hat{H}t} e^{-i\hat{\vec{P}} \cdot \vec{x}}\chi_i(0)
  e^{i\hat{\vec{P}} \cdot \vec{x}} e^{-\hat{H}t}
\right| B ,p',s \right\rangle \nonumber \\
&&\hspace*{4cm} \times \left\langle B ,p',s \left| \overline{\chi}_j(0)
\right| \Omega \right\rangle                            \nonumber \\
&=& \sum_{s} \sum_{B} e^{-{E_{B}t}}
\left\langle\Omega| \chi_i(0) |B ,p,s \rangle
 \langle B ,p,s | \overline{\chi}_j(0)|\Omega\right\rangle\ .
\end{eqnarray}
%

It is convenient in the following discussion to label the states
which have the $\chi$ interpolating field quantum numbers and
which survive the parity projection as $|B_{\alpha}\rangle$ for
$\alpha=1,2,\cdots,N$.  In general the number of states,
$N$, in this tower of excited states may be infinite, but we
will only ever need to consider a finite set of the lowest such
states here.
After selecting zero momentum, $\vec p=0$, the parity-projected trace
of this object is then
\begin{equation}
\label{eqn:Gijequation}
G^{\pm}_{ij}(t)
 = {\rm tr}[\Gamma_{\pm} {\cal G}_{ij}(t,\vec 0)]
 = \sum_{\alpha=1}^{N} e^{-{m_{\alpha}}t}
   \lambda^{\alpha}_i \overline{\lambda}^{\alpha}_j\ ,
\end{equation}
where $\lambda^{\alpha}_i$ and $\overline{\lambda}^{\alpha}_j$ are
coefficients denoting the
couplings of the interpolating fields $\chi_i$ and $\overline{\chi}_j$,
respectively, to the state $\left|B_{\alpha}\right\rangle$.
If we use identical source and sink interpolating fields then it
follows from the definition of the coupling strength that
$\overline{\lambda}^{\alpha}_j = (\lambda^{\alpha}_j)^*$
and from Eq.~(\ref{eqn:Gijequation}) we see that
$G^{\pm}_{ij}(t)=[G^{\pm}_{ji}(t)]^*$, i.e., $G^\pm$ is
a Hermitian matrix.  If, in addition, we use only real
coefficients in the link products, then $G^\pm$ is a real
symmetric matrix.  For the correlation matrices that we
construct, we have real link coefficients but smeared
sources and point sinks.
Consequently, $G$ is a real but non-symmetric matrix.
Since $G^\pm$ is a real matrix for the
infinite number of possible choices of interpolating fields with
real coefficients, then we can take
$\lambda^{\alpha}_i$ and $\overline{\lambda}^{\alpha}_j$ to
be real coefficients without loss of generality.

Suppose now that we have $M$ creation and annihilation operators,
where $M < N$.
We can then form an $M \times M$ approximation to the full $N \times N$
matrix $G$.
At this point there are two options for extracting masses.
The first is the standard method for calculation of effective masses
at large $t$ described in Section~\ref{spin12LT}.
%
%
The second option is to extract the masses through a correlation-matrix
procedure~\cite{MCNEILE}.

Let us begin by considering the ideal case where we have $N$
interpolating fields with the same quantum numbers, but which
give rise to $N$ linearly independent states when acting on the
vacuum.  In this case we can construct $N$ ideal interpolating
source and sink fields which perfectly isolate the $N$ individual
baryon states $|B_\alpha\rangle$,
%
\begin{eqnarray}
\overline{\phi}^{\alpha} &=& \sum_{i=1}^N 
          u^{\alpha}_i\ \overline{\chi}_i\ , \\
\phi^{\alpha} &=& \sum_{i=1}^N v^{\ast \alpha}_i\ \chi_i\ ,
\label{lincomIF}
\end{eqnarray}
%
such that
%
\begin{eqnarray}
\left\langle B_{\beta}\right| \overline{\phi}^{\alpha}
\left| \Omega\right\rangle
&=& \delta_{\alpha\beta}\ \overline{z}^{\alpha}\
\overline{u}(\alpha,p,s)\ ,                     \\
\left\langle \Omega \right | \phi^{\alpha}
\left| B_{\beta}\right\rangle
&=& \delta_{\alpha\beta}\ z^{\alpha}\
u(\alpha,p,s)\ ,
\label{PhiExpression}
\end{eqnarray}
%
where $z^\alpha$ and $\overline{z}^\alpha$ are the coupling strengths
of $\phi^\alpha$ and $\overline{\phi}^\alpha$ to the state
$|B_\alpha\rangle$.
The coefficients $u_i^\alpha$ and $v_i^{\ast \alpha}$ in
Eqs.~(\ref{lincomIF}) may differ when the source and sink have different
smearing prescriptions, again indicated by the differentiation between
$z^\alpha$ and $\overline{z}^\alpha$.  
For notational convenience for the remainder of this
discussion repeated indices $i,j,k$ are to be understood
as being summed over. 
At $\vec{p}=0$, it follows that,
\begin{eqnarray}
\label{ActingLeft}
G^{\pm}_{ij}(t)\ u^{\alpha}_j
&=& \left(\sum_{\vec{x}} {\rm tr}
\left[ \Gamma_{\pm}
  \left \langle \Omega \right |
  \chi_i \overline{\chi}_j
  \left| \Omega \right\rangle
\right]
\right) u^{\alpha}_j                    \nonumber\\
&=& \lambda^{\alpha}_i \overline{z}^{\alpha} 
    e^{-m_{\alpha} t} .
\end{eqnarray}
The only $t$-dependence in this expression comes from the exponential
term, which leads to the recurrence relationship
\begin{equation}
G^{\pm}_{ij} (t) \, u^{\alpha}_j
= e^{m_{\alpha}} G^{\pm}_{ik} (t+1) \, u^{\alpha}_k\ ,
\label{eveqn}
\end{equation}
which can be rewritten as
\begin{equation}
[G^{\pm} (t+1)]_{ki}^{-1} G^{\pm}_{ij} (t) \, u^{\alpha}_j
= e^{m_{\alpha}}\, u^{\alpha}_k\ .
\label{eveqn2}
\end{equation}
This is recognised as an eigenvalue equation for the matrix
$[G^{\pm} (t+1)]^{-1} G^{\pm}(t)$ with 
eigenvalues $e^{m_{\alpha}}$ and eigenvectors $u^\alpha$.
Hence the natural logarithms of the eigenvalues of
$[G^{\pm} (t+1)]^{-1} G^{\pm}(t)$ are the masses of the
$N$ baryons in the tower of excited states corresponding
to the selected parity and the quantum numbers of the
$\chi$ fields.  The eigenvectors are the coefficients of the
$\chi$ fields providing the ideal linear combination for that
state.  Note that since here we use only real coefficients in our
link products, then $[G^{\pm} (t+1)]^{-1} G^{\pm}(t)$ is a real
matrix and so $u^\alpha$ and $v^\alpha$ will be real eigenvectors.
It also then follows that $z^{\alpha}$ and $\overline{z}^\alpha$
will be real.
These coefficients are examined in detail in the following
section.

One can also construct the equivalent left-eigenvalue equation to recover
the $v$ vectors, providing the optimal linear combination of
annihilation interpolators,
\begin{equation}
v^{\ast \alpha}_k G^{\pm}_{kj} (t)
= e^{m_{\alpha}} v^{\ast \alpha}_i G^{\pm}_{ij} (t+1)\ .
\end{equation}
Recalling Eq.~(\ref{ActingLeft}), one finds:
\begin{eqnarray}
G^{\pm}_{ij} (t)\ u^{\alpha}_j
&=& \overline{z}^{\alpha} \lambda^{\alpha}_i
    e^{-m_{\alpha} t}\ ,                                   \\
v^{\ast \alpha}_i\ G^{\pm}_{ij} (t)
&=& {z}^{\alpha} \overline{\lambda}^{\alpha}_j e^{-m_{\alpha} t }\ , \\
v^{\ast \alpha}_k\ G^{\pm}_{kj} (t) G^{\pm}_{il} (t)\ u^{\alpha}_l
&=& z^{\alpha} \overline{z}^{\alpha} \lambda^{\alpha}_i 
\overline{\lambda}^{\alpha}_j e^{-2m_{\alpha} t}\ .
\label{preprojection}
\end{eqnarray}
The definitions of Eqs.~(\ref{PhiExpression}) imply
\begin{equation}
v^{\ast \alpha}_i\ G^{\pm}_{ij}(t)\ u^{\alpha}_j =
z^{\alpha}\overline{z}^{\alpha} e^{-m_{\alpha} t } ,
\label{projection}
\end{equation}
indicating the eigenvectors may be used to construct a correlation
function in which a single state mass $m_\alpha$is isolated
and which can be analysed
using the methods of Section~II. We refer to this as the projected
correlation function in the following.
Combining Eqs.~(\ref{preprojection}) and (\ref{projection}) leads us to
the result
\begin{equation}
\frac{v^{\ast \alpha}_k \ G_{kj}(t) G_{il}(t)\ u^{\alpha}_l}
{v^{\ast \alpha}_k G_{kl}(t) u^{\alpha}_l }
= \lambda^{\alpha}_{i}\overline{\lambda}^{\alpha}_{j}
  e^{-m_{\alpha} t} 
\ .
\label{eqn:ratios}
\end{equation}
By extracting all $N^2$ such ratios, we can exactly
recover all of the real couplings $\lambda^{\alpha}_{i}$ and 
$\overline{\lambda}^{\alpha}_{j}$ of 
$\chi_i$ and $\overline{\chi}_j$ respectively to
the state $|B_\alpha\rangle$.
Note that throughout this section no assumptions have been made about
the symmetry properties of $G_{ij}^{\pm}$. This is essential due to our
use of smeared sources and point sinks.

In practice we will only have a relatively small number, $M<N$,
of interpolating fields in any given analysis.  These $M$
interpolators should be chosen to have good overlap with the
lowest $M$ excited states in the tower and we should attempt
to study the ratios in Eq.~(\ref{eqn:ratios}) at early to
intermediate Euclidean times, where the contribution of the
$(N-M)$ higher mass states will be suppressed but where
there is still sufficient signal to allow the lowest $M$ states
to be seen.  This procedure will lead
to an estimate for the masses of each of the lowest $M$ states
in the tower of excited states.  Of these $M$
predicted masses, the highest will in general have the largest
systematic error while the lower masses will be
most reliably determined.  Repeating the analysis with varying
$M$ and different combinations of interpolating fields
will give an objective measure of the reliability of the
extraction of these masses.

In our case of a modest $2 \times 2$ correlation matrix ($M=2$)
we take a cautious approach to the selection of the
eigenvalue analysis time.
As already explained, we perform the eigenvalue analysis at an early
to moderate Euclidean time where statistical noise is suppressed
and yet contributions from at least the lowest two mass states
is still present.  One must exercise caution in performing the
analysis at too early a time, as more than the desired
$M=2$ states may be contributing to the $2\times 2$ matrix of
correlation functions.

We begin by projecting a particular parity, and then investigate the
effective mass plots of the elements of the correlation matrix.  Using
the covariance-matrix based $\chi^2/N_{\rm DF}$, we identify the time
slice at which all correlation functions of the correlation matrix are
dominated by a single state.  In practice, this time slice is
determined by the correlator providing the lowest-lying effective mass
plot. 
The eigenvalue analysis is performed at one time slice earlier,
thus ensuring the presence of multiple states in the elements of the
correlation function matrix, minimising statistical uncertainties,
and hopefully providing a clear signal for the analysis.  
In this approach minimal new
information has been added, providing the best opportunity that the $2
\times 2$ correlation matrix is indeed dominated by 2 states.  The
left and right eigenvectors are determined and used to project
correlation functions containing a single state from the correlation
matrix as indicated in Eq.~(\ref{projection}).  These correlation
functions are then subjected to the same covariance-matrix based
$\chi^2/N_{\rm DF}$ analysis to identify new acceptable fit windows for
determining the masses of the resonances.


%

\end{document}